\documentclass[aps,prb,twocolumn,groupedaddress,amsfonts,showpacs]{revtex4}
\usepackage{graphicx,color}
\usepackage{bm}

\begin{document}
\title{Anisotropic tunneling
magnetoresistance and tunneling anisotropic magnetoresistance:
spin-orbit coupling in magnetic tunnel junctions}
\author{A. Matos-Abiague and J. Fabian}
\affiliation{Institute for Theoretical Physics, University of
Regensburg, 93040 Regensburg, Germany}
\date{\today}

\begin{abstract}
The effects of the spin-orbit interaction on the tunneling
magnetoresistance of ferromagnet/semiconductor/normal metal tunnel
junctions are investigated. Analytical expressions for the tunneling
anisotropic magnetoresistance (TAMR) are derived within an
approximation in which the dependence of the magnetoresistance on
the magnetization orientation in the ferromagnet originates from the
interference between Bychkov-Rashba and Dresselhaus spin-orbit
couplings that appear at junction interfaces and in the tunneling
region. We also investigate the transport properties of
ferromagnet/semiconductor/ferromagnet tunnel junctions and show that
in such structures the spin-orbit interaction leads not only to the
TAMR effect but also to the anisotropy of the conventional tunneling
magnetoresistance (TMR). The resulting anisotropic tunneling
magnetoresistance (ATMR) depends on the absolute magnetization
directions in the ferromagnets. Within the proposed model, depending
on the magnetization directions in the ferromagnets, the interplay
of Bychkov-Rashba and Dresselhaus spin-orbit couplings produces
differences between the rates of transmitted and reflected spins at
the ferromagnet/seminconductor interfaces, which results in an
anisotropic local density of states at the Fermi surface and in the
TAMR and ATMR effects. Model calculations for Fe/GaAs/Fe tunnel
junctions are presented. Furthermore, based on rather general
symmetry considerations, we deduce the form of the magnetoresistance
dependence on the absolute orientations of the magnetizations in the
ferromagnets.
\end{abstract}

\pacs{73.43.Jn, 72.25.Dc, 73.43.Qt} \keywords{TAMR, tunneling
magnetoresistance, ferromagnte/semiconductor/ferromagnet tunnel
junctions, spin-orbit coupling}

\maketitle

\section{Introduction}
\label{intro}

The tunneling magnetoresistance (TMR) effect is observed in
ferromagnet/insulator/ferromagnet heterojunctions, in which the
magnetoresistance exhibits a strong dependence on the relative
magnetization directions in the two ferromagnetic layers and on
their spin
polarizations.\cite{Julliere1975:PL,Maekawa:2002,Slonczewski1989:PRB,Zutic2004:RMP,Fabian2007:APS}
Because of this peculiarly strong asymmetric behavior of the
magnetoresistance, TMR devices find multiple uses ranging from
magnetic sensors to magnetic random access memory
applications.\cite{Maekawa:2002,Zutic2004:RMP}

Beyond the conventional TMR effect, it has been observed that the
magnetoresistance in magnetic tunnel junctions (MTJs) may also
depend on the {\it absolute} orientation of the magnetizations in
the ferromagnetic
leads.\cite{Gould2004:PRL,Ruster2005:PRL,Saito2005:PRL,Brey2004:APL}
This phenomenon is called the tunneling anisotropic
magnetoresistance (TAMR) effect.\cite{Gould2004:PRL,Brey2004:APL} A
theoretical investigation of the tunneling magnetoresistance in
GaMnAs/GaAlAs/GaMnAs tunnel junctions was reported in Ref.~
\onlinecite{Brey2004:APL} which predicted that, as a result of the
strong spin-orbit interaction the tunneling magnetoresistance
depends on the angle between the current flow direction and the
orientation of the electrode magnetization. A difference between the
tunneling magnetoresistances in the in-plane (i.e., magnetization in
the plane of the magnetic layers) and out-of-plane configurations of
up to $6\%$ was predicted for large values of the electrode spin
polarization.\cite{Brey2004:APL} Here we refer to this phenomenon as
the \emph{out-of-plane} TAMR. Recent first principles calculations
in Fe/MgO/Fe magnetic tunnel junctions (MTJs) predict an
out-of-plane TAMR ratio of about 44 \%.\cite{Khan2008:JPCM} On the
other hand, we refer to an \emph{in-plane} TAMR effect as the change
in the magnetoresistance when the in-plane magnetization of the
ferromagnetic layer(s) is rotated in the plane perpendicular to the
direction of the current flow.

It is remarkable that the TAMR is present even in MTJs in which only
one of the electrodes is magnetic and the conventional TMR is
absent. In contrast to the conventional TMR-based devices, which
require two magnetic layers for their operation, TAMR-based devices
can operate with a single magnetic lead, opening new possibilities
and functionalities for the operation of spintronic devices. The
TAMR may also affect the spin-injection from a ferromagnet into a
non-magnetic semiconductor. Therefore, in order to correctly
interpret the results of spin injection experiments in a spin-valve
configuration, it is essential to understand the nature, properties,
and origin of the TAMR effect.

The first experimental observation of TAMR was in (Ga,Mn)As/AlOx/Au
heterojunctions, in which an in-plane TAMR ratio of about $2.7\%$
was found.\cite{Gould2004:PRL} Experimental investigations of the
in-plane TAMR in (Ga,Mn)As/GaAs/(Ga,Mn)As and in
(Ga,Mn)As/ZnSe/(Ga,Mn)As tunnel junctions, in which both electrodes
are ferromagnetic have also been
reported.\cite{Ruster2005:PRL,Saito2005:PRL} In the case of
(Ga,Mn)As/ZnSe/(Ga,Mn)As, the in-plane TAMR ratio was found to
decrease with increasing temperature, from about $10\%$ at 2 K to
$8.5\%$ at 20 K.\cite{Saito2005:PRL} This temperature dependence of
the in-plane TAMR is more dramatic in the case of
(Ga,Mn)As/GaAs/(Ga,Mn)As, for which a TAMR ratio of the order of a
few hundred percent at 4 K was amplified to $150\;000\%$ at 1.7
K.\cite{Ruster2005:PRL} This huge amplification of the in-plane TAMR
was suggested to originate from the opening of the Efros-Shklovskii
gap\cite{Efros1975:JPC} at the Fermi energy when crossing the
metal-insulator transition.\cite{Ruster2005:PRL} Measurements of the
TAMR in $\textrm{p}^{+}-$(Ga,Mn)As/$\textrm{n}^{+}$-GaAs Esaki diode
devices have also been reported.\cite{Ciorga2007:NJP} In addition to
the investigations involving vertical tunneling devices the TAMR has
also been studied in break
junctions,\cite{Bolotin2006:PRL,Burton2007:PRB}
nanoconstrictions\cite{Giddings2005:PRL,Ciorga2007:NJP} and
nanocontacts.\cite{Jacob2008:PRB}

Beyond the area of currently low Curie temperature ferromagnetic
semiconductors, the TAMR has recently been experimentally
investigated in Fe(001)/vacuum/bcc-Cu(001) tunnel junctions,
\cite{Chantis2007:PRL} Fe/GaAs/Au
MTJs,\cite{Moser2007:PRL,Fabian2007:APS} Co/AlO$_x$/Au
MTJs,\cite{Liu2008:NL}, CoPt structures\cite{Shick2006:PRB} and in
multilayer-(Co/Pt)/AlO$_x$/Pt structures.\cite{Park2008:PRL}

In what follows we focus our discussion on the case of the in-plane
TAMR (for brevity we will refer to it as the TAMR effect). We
investigate the TAMR in ferromagnet/semiconductor/normal metal
(F/S/NM) and in ferromagnet/semiconductor/ferromagnet (F/S/F) MTJs.
We propose a model in which the two-fold symmetric magnetoresistance
dependence on the orientation of the in-plane magnetization in the
ferromagnetic layer(s) originates from the interference of
Dresselhaus and Bychkov-Rashba-like spin-orbit couplings. Such
interference effects have already been investigated in lateral
transport in 2D electron
systems\cite{Trushin2007:PRB,Cheng2007:PRB,Chalaev2008:PRB}, in spin
relaxation in quantum wells \cite{Averkiev1999:PRB} and quantum dots
\cite{Stano2006:PRL}, or in 2D plasmons.\cite{Badalyan2008:arXiv}
The symmetry, which is imprinted in the tunneling probability
becomes apparent when a magnetic moment is present. Our main results
are: i) finding analytical expressions for evaluating the TAMR in
both F/S/NM and F/S/F MTJs, ii) prediction and evaluation of the
ATMR in F/S/F heterojunctions, and iii) derivation of a simple
phenomenological relation describing the dependence of the tunneling
magnetoresistance on the absolute orientation of the
magnetization(s) of the ferromagnet(s).

The paper is organized as follows. In Sec.~\ref{theo} we present the
theoretical model describing the tunneling through a MTJ. In a first
approximation we consider the case of an infinitesimally thin
barrier (Sec.~\ref{ddm}), while the finite spatial extension of the
potential barrier is incorporated in a more sophisticated approach
discussed in Sec.~\ref{ssom}. Detailed solutions and tunneling
properties within these approximations are given in Appendices
\ref{appA} and \ref{appB}, respectively. In Sec.~\ref{tamr-sec} we
discuss the TAMR in both F/S/NM (Sec.~\ref{tamr-fsn}) and F/S/F
(Sec.~\ref{tamr-fsf}) MTJs. The ATMR in F/S/F tunnel junctions is
investigated in Sec.~\ref{atmr-sec}, where specific calculations for
model Fe/GaAs/Fe MTJs are presented. In Sec.~\ref{pheno} we develop
a phenological model for determining the dependence of the TAMR and
ATMR on the absolute orientation(s) of the magnetization(s) in the
ferromagnetic lead(s). Finally, conclusions are given in
Sec.~\ref{conclu}.

\section{Theoretical model.}
\label{theo}

Consider a F/S/F tunnel heterojunction. The semiconductor is
assumed to lack bulk inversion symmetry (zinc-blende
semiconductors are typical examples). The bulk inversion asymmetry
of the semiconductor together with the structure inversion
asymmetry (for the case of asymmetric junctions) of the
heterojunction give rise to the
Dresselhaus\cite{Dresselhaus1955:PR,Rossler2002:SSC,Winkler:2003,Fabian2007:APS}
and
Bychkov-Rashba\cite{Bychkov1984:JPC,Winkler:2003,Fabian2007:APS}
SOIs, respectively. The interference of these two spin-orbit
couplings leads to a net anisotropic SOI with a $C_{2v}$ symmetry
which is imprinted onto the tunneling magnetoresistance as the
electrons pass through the semiconductor barrier. This was
discussed in some details in
Refs.~\onlinecite{Moser2007:PRL,Fabian2007:APS} for the case of
F/S/NM tunnel junctions. Here we generalize the model proposed in
Refs.~\onlinecite{Moser2007:PRL,Fabian2007:APS} to the case of
F/S/F tunnel junctions. For such structures our model predicts the
coexistence of both the TAMR and ATMR phenomena.

We consider a F/S/F tunnel junction grown in the $z=[001]$
direction, where the semiconductor forms a barrier of width $d$
between the left and right ferromagnetic electrodes. At first we
discuss a simplified model for very thin barriers. In that case
the barrier can be approximated by a Dirac delta function and the
SOI reduced to the plane of the barrier. In what follows we will
refer to this model as the Dirac delta model (DDM). A second model
in which Slonczewski's
proposal\cite{Slonczewski1989:PRB,Fabian2007:APS} for
ferromagnet/insulator/ferromagnet tunnel junctions is generalized
to the case of ferromagnet/semiconductor/ferromagnet junctions by
including the Bychkov-Rashba and Dresselhaus SOIs will be referred
to as the Slonczewski spin-orbit model (SSOM).

\subsection{Dirac delta model (DDM)}
\label{ddm}

We consider here the case of a very thin tunneling barrier.
Assuming that the in-plane wave vector $\mathbf{k}_{\parallel}$ is
conserved throughout the heterostructure, one can decouple the
motion along the growth direction ($z$) from the other spatial
degrees of freedom. The effective model Hamiltonian describing the
tunneling across the heterojunction reads
\begin{equation}\label{hamilt}
    H=H_{0}+H_{Z}+H_{SO}.
\end{equation}
Here
\begin{equation}\label{h0}
H_{0}=-\frac{\hbar^2}{2m_{0}}\frac{d^2}{dz^2}+V_{0}d\delta(z),
\end{equation}
with $m_{0}$ the bare electron mass, and $V_{0}$ and $d$ the high
and width, respectively, of the actual potential barrier [here
modelled with a Dirac delta function $\delta(z)$] along the growth
direction ($z=[001]$) of the heterostructure.

The spin splitting due to the exchange field in the left $(z<0)$
and right $(z>0)$ ferromagnetic regions is given by
\begin{equation}\label{zeeman}
    H_{Z}=-\frac{\Theta(-z)\Delta_{l}}{2}
    \mathbf{n}_{l}\cdot \mbox{\boldmath$\sigma$}-\frac{\Theta(z)\Delta_{r}}{2}
    \mathbf{n}_{r}\cdot \mbox{\boldmath$\sigma$}.
\end{equation}
Here $\Delta_{l}$ and $\Delta_{r}$ represent the exchange energy
in the left and right ferromagnets, respectively, and $\Theta(z)$
is the Heaviside step function. The components of the vector
$\mbox{\boldmath$\sigma$}$ are the Pauli matrices, and
$\mathbf{n}_{j}=(\cos\theta_{j},\sin\theta_{j},0)$ with $j=l,r$ is
a unit vector defining the in-plane magnetization direction in the
left ($j=l$) and right ($j=r$) ferromagnets with respect to the
[100] crystallographic direction. The Zeeman splitting in the
semiconductor can be neglected.

We note that in recent experiments with Fe/GaAs/Au tunnel
junctions,\cite{Moser2007:PRL} the reference axis was taken as the
[110] direction.
%, which corresponds to the hard axis of
%magnetization in the Fe layer.
Therefore, it is convenient to express the magnetization direction
relative to the [110] axis by introducing the angle shifting
$\phi_{j}=\theta_{j} - \pi/4$ ($j=l,r$). One can then write
$\mathbf{n}_{j}=[\cos(\phi_{j}+\pi/4),\sin(\phi_{j}+\pi/4),0]$
with $\phi_{j}$ giving the magnetization direction in the left
($j=l$) and right ($j=r$) ferromagnets with respect to the [110]
crystallographic direction.

Within the DDM, the spin-orbit interaction throughout the
semiconductor barrier (including the interfaces) can be written as
\begin{equation}\label{so-delta}
    H_{SO}=(\mathbf{w}\cdot \mbox{\boldmath$\sigma$})\delta(z),
\end{equation}
with the effective spin-orbit coupling field
\begin{equation}\label{rashba-delta}
   \mathbf{w}=(-\bar{\alpha}k_{y}+\bar{\gamma}k_{x},\bar{\alpha}k_{x}-\bar{\gamma}k_{y},0).
\end{equation}
Here $\bar{\alpha}$ and $\bar{\gamma}$ represent
% the average
effective values of the Bychkov-Rashba and linearized Dresselhaus
parameters, respectively, and $k_{x}$ and $k_{y}$ refer to the $x$
and $y$ components of the wave vector $\mathbf{k}$. In terms of
the usual Dresselhaus parameter $\gamma$, the linearized
Dresselhaus parameter can be approximated\cite{Fabian2007:APS} as
$\bar{\gamma}\approx \gamma Q$, where $Q=2m_{0}V_{0}d/\hbar^2$
stands for the strength of the effective wave vector in the
barrier.

The scattering states in the left ($z<0$) and right ($z>0$)
ferromagnetic regions are given by
\begin{equation}\label{scattL}
\Psi_{\sigma}^{(l)}=\frac{e^{ik_{\sigma}z}\chi_{\sigma}^{(l)}}{\sqrt{k_{\sigma}}
}+
    r_{\sigma,\sigma}e^{-ik_{\sigma}z}\chi_{\sigma}^{(l)}+
    r_{\sigma,-\sigma}e^{-ik_{-\sigma}z}\chi_{-\sigma}^{(l)},
\end{equation}
and
\begin{equation}\label{scattR}
    \Psi_{\sigma}^{(r)}=
    t_{\sigma,\sigma}e^{i\kappa_{\sigma}z}\chi_{\sigma}^{(r)}+
    t_{\sigma,-\sigma}e^{i\kappa_{-\sigma}z}\chi_{-\sigma}^{(r)},
\end{equation}
respectively. Here we have introduced the wave vector components
\begin{equation}\label{kvect}
    k_{\sigma}=\sqrt{\frac{2m_{0}}{\hbar^2}\left(E+\sigma
    \frac{\Delta_{l}}{2}\right)-k_{\parallel}^{2}},
\end{equation}
and
\begin{equation}\label{kapvect}
    \kappa_{\sigma}=\sqrt{\frac{2m_{0}}{\hbar^2}\left(E+\sigma
    \frac{\Delta_{r}}{2}\right)-k_{\parallel}^{2}},
\end{equation}
with $k_{\parallel}=\sqrt{k_{x}^{2}+k_{y}^{2}}$ denoting the
length of the wave vector component corresponding to the free
motion in the $x-y$ plane.
The spinors
\begin{equation}\label{spin-L}
    \chi_{\sigma}^{(j)}=\frac{1}{\sqrt{2}}\left(%
\begin{array}{c}
  1 \\
  \sigma e^{i(\phi_{j}+\pi/4)} \\
\end{array}%
\right)\;\;\;(j=l,r),
\end{equation}
correspond to a spin parallel ($\sigma = \uparrow$) or
antiparallel ($\sigma = \downarrow$) to the magnetization
direction
$\mathbf{n}_{j}=[\cos(\phi_{j}+\pi/4),\sin(\phi_{j}+\pi/4),0]$ in
the left ($j=l$) and right ($j=r$) ferromagnets.

The reflection and transmission coefficients can be found by
imposing appropriate boundary conditions and solving the
corresponding system of linear equations (for details see Appendix
\ref{appA}). The transmissivity of an incoming spin-$\sigma$
particle can then be evaluated from the relation
\begin{equation}\label{trans}
T_{\sigma}(E,k_{\parallel})=\textrm{Re}[\kappa_{\sigma}(|t_{\sigma,\sigma}|
^2+\kappa_{-\sigma}|t_{\sigma,-\sigma}|^2)].
\end{equation}
Explicit analytical expressions for the transmission coefficients
$(t_{\sigma,\sigma}$ and $t_{\sigma,-\sigma})$ are given in
Appendix \ref{appA}.

\subsection{Slonczewski spin-orbit model (SSOM)}
\label{ssom}

We give a generalization of the Slonczewski
model\cite{Slonczewski1989:PRB} for
ferromagnet/insulator/ferromagnet tunnel junctions to the case in
which the insulator barrier is replaced by a zinc-blende
semiconductor. Unlike in the DDM, now the spatial extension of the
potential barrier is taken into account. The model Hamiltonian is
\begin{equation}\label{hamilt-ssom}
    H=H_{0}+H_{Z}+H_{BR}+H_{D},
\end{equation}
where
\begin{equation}\label{h0-ssom}
H_{0}=-\frac{\hbar^2}{2}\nabla\left[\frac{1}{m(z)}\nabla\right]+V_{0}\Theta(z)\Theta(d-z).
\end{equation}
The electron effective mass $m(z)$ is assumed to be $m = m_{c}$ in
the central (semiconductor) region and $m=m_{0}$ in the
ferromagnets. The exchange splitting in the ferromagnets is now
given by
\begin{equation}\label{zeeman-ssom}
    H_{Z}=-\frac{\Theta(-z)\Delta_{l}}{2}
    \mathbf{n}_{l}\cdot \mbox{\boldmath$\sigma$}-\frac{\Theta(z-d)\Delta_{r}}{2}
    \mathbf{n}_{r}\cdot \mbox{\boldmath$\sigma$}.
\end{equation}

The Dresselhaus SOI can be written as\cite{Rossler2002:SSC,
Winkler:2003,Ganichev2004:PRL,Zawadzki2004:SST,Fabian2007:APS}
\begin{equation}\label{dresselhaus}
    H_{D}=(k_{x}\sigma_{x}-k_{y}\sigma_{y})
    \frac{\partial}{\partial z}\left(\gamma(z)\frac{\partial}{\partial
    z}\right),
\end{equation}
where $x$ and $y$ correspond to the $[100]$ and $[010]$
directions, respectively. The Dresselhaus parameter $\gamma(z)$
has a finite value $\gamma$ in the semiconductor region, where the
bulk inversion asymmetry is present, and vanishes elsewhere. Note
that because of the step-like spatial dependence of $\gamma(z)$,
the Dresselhaus SOI [Eq.~(\ref{dresselhaus})] implicitly includes
both the interface and bulk
contributions.\cite{Rossler2002:SSC,Zawadzki2004:SST,Fabian2007:APS}

The Bychkov-Rashba SOI is given by\cite{note1}
\begin{equation}\label{rashba-inter}
H_{BR}
%^{\textrm{\small{interface}}}
=\left[\alpha_{l}\delta(z-z_{l})-\alpha_{r}\delta(z-z_{r})\right]
(k_{x}\sigma_{y}-k_{y}\sigma_{x}),
\end{equation}
and arises due to the ferromagnet/semiconductor interface
inversion asymmetry.\cite{Fabian2007:APS} Here $\alpha_l$
($\alpha_r$) denotes the SOI strength at the left (right)
interface $z_{l}=0$ ($z_{r}=d$). For the small voltages considered
here (up to a hundred mV), the Bychkov-Rashba SOI inside the
semiconductor can be neglected.

The $z$-components of the scattering states in the left and right
ferromagnets have the same form as in Eqs.~(\ref{scattL}) and
(\ref{scattR}), respectively.

In the central (semiconductor) region ($0<z<d$) we have
\begin{equation}\label{scattC}
    \Psi_{\sigma}^{(c)}=
\sum_{i=\pm}(A_{\sigma,i}e^{q_{i}z}+B_{\sigma,i}e^{-q_{i}z})\chi_{i}^{(c)},
\end{equation}
\begin{equation}\label{spin-Ln}
    \chi_{\pm}^{(c)}=\frac{1}{\sqrt{2}}\left(%
\begin{array}{c}
  1 \\
  \pm e^{i\xi} \\
\end{array}%
\right).
\end{equation}
The angle $\xi$ is defined through the relation
$\tan(\xi)=-k_{y}/k_{x}$. We have also used the notation
\begin{equation}\label{q-def}
    q_{\pm}=\frac{q_{0}}{\sqrt{1\mp \left(\frac{2m_{c}\gamma
k_{\parallel}}{\hbar^2}\right)^{2}}},
\end{equation}
where
\begin{equation}\label{q0-def}
    q_{0}=\sqrt{\frac{2m_{c}(V_{0}-E)}{\hbar^{2}}+k_{\parallel}^{2}},
\end{equation}
is the length of the $z$ component of the wave vector in the
barrier in the absence of SOI.

The expansion coefficients in Eqs.~(\ref{scattL}), (\ref{scattR}),
and (\ref{scattC} can be found by applying appropriate matching
conditions at each interface and by solving the corresponding
system of linear equations (for details, see Appendix \ref{appB}).
Once the wave function is determined, the particle transmissivity
can be calculated from Eq.~(\ref{trans}). Approximate analytical
expressions for the transmission coefficients $t_{\sigma,\sigma}$
and $t_{\sigma,-\sigma}$ are given in Appendix \ref{appB}.

\section{Tunneling anisotropic magnetoresistance (TAMR)}
\label{tamr-sec}

The magnetoresistance of a tunnel junction can be obtained by
evaluating the current through the device or the conductance. The
current flowing along the heterojunction is given by
\begin{equation}\label{current}
    I=\frac{e}{(2\pi)^3\hbar}\sum_{\sigma =\uparrow,\downarrow}\int
    dE
d^2\mathbf{k}_{\parallel}T_{\sigma}(E,\mathbf{k}_{\parallel})[f_{l}(E)-f_{r}(E)],
\end{equation}
where $f_l(E)$ and $f_{r}(E)$ are Fermi-Dirac distributions with
chemical potentials $\mu_{l}$ and $\mu_{r}$ in the left
%(ferromagnetic)
and right
(metallic or ferromagnetic)
leads,
respectively. For the case of zero temperature and small voltages,
the Fermi-Dirac distributions can be expanded in powers of the
voltage $V_{bias}$. To first order in $V_{bias}$ one obtains
$f_{l}(E)-f_{r}(E)\approx \delta(E-E_{F})V_{bias}$ with
$\delta(x)$ the Dirac delta function and $E_{F}$ the Fermi energy.
One then obtains the following approximate expression for the
conductance
\begin{equation}\label{conductance}
    G=\sum_{\sigma =\uparrow,\downarrow}G_{\sigma}\;,\;\;
    G_{\sigma}=\frac{e^2}{(2\pi)^3 \hbar}\int
    d^2\mathbf{k}_{\parallel}T_{\sigma}(E_{F},\mathbf{k}_{\parallel}).
\end{equation}
We note that although similar, the expression above differs from
the linear response conductance. In our case, the transmissivity
$T_{\sigma}(E_{F},\mathbf{k}_{\parallel})$ depends on the
Bychkov-Rashba parameter $\bar{\alpha}$. Recent first-principles
calculations\cite{Gmitra2008} have shown that the spin-orbit
coupling field is different for different bands, therefore the
effective value of $\bar{\alpha}$ is energy dependent. By applying
an external voltage the energy window relevant for tunneling can
be changed, resulting in voltage-dependent values of
$\bar{\alpha}$. Consequently, the conductance in
Eq.~(\ref{conductance}) depends, parametrically, on the applied
voltage.

\subsection{TAMR in ferromagnet/semiconductor/normal metal tunnel
junctions}
\label{tamr-fsn}

The tunneling properties of F/S/NM junctions can be obtained as a
limit case of the models proposed in Sec. \ref{theo} for F/S/F
tunnel junctions by taking $\phi_{l}=\phi_{r}=\phi$ and
$\Delta_{r}$ as the Zeeman splitting in the normal metal region.
In the present case $l$, $c$, and $r$ refer to the ferromagnetic
(left), semiconductor (central), and normal metal (right) regions,
respectively.

\begin{figure}
\includegraphics[width=7cm]{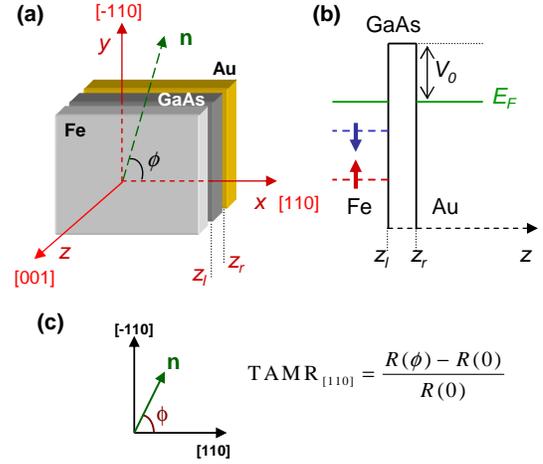}
\caption{(color online). (a) Schematics of a Fe/GaAs/Au MTJ. The
magnetization direction in the ferromagnet is specified by the
vector $\mathbf{n}$. (b) Schematics of the potential profile of
the heterojunction along the [001] direction. (c) Definition of
the TAMR in F/S/NM junctions.} \label{fig-fsn}
\end{figure}

The TAMR in F/S/NM tunnel junctions refers to the changes of the
tunneling magnetoresistance ($R$) when varying the magnetization
direction $\mathbf{n}_{l}$ of the magnetic layer with respect to a
fixed axis. Here we assume the [110] crystallographic direction as
the reference axis. The TAMR is then given by\cite{Fabian2007:APS}
\begin{equation}\label{tamr-def}
    \textrm{TAMR}_{[110]}(\phi)=\frac{R(\phi)-R(0)}{R(0)}=
    \frac{G(0)-G(\phi)}{G(\phi)}.
\end{equation}
Since in a F/S/NM tunnel junction only one electrode is magnetic,
the conventional TMR effect is absent.

An alternative to the magnetoresistance, which refers to the
charge transport, is the spin polarization efficiency of the
transmission characterized by the tunneling spin
polarization\cite{Maekawa:2002,Zutic2004:RMP}
\begin{equation}\label{s-pol}
    P=\frac{I_{\uparrow}-I_{\downarrow}}{I},
\end{equation}
where $I_{\sigma}$ is the charge current corresponding to the
spin-$\sigma$ channel and $I$ is the total current. The changes in
the tunneling spin polarization when the magnetization of the
ferromagnet is rotated in-plane can then be characterized by the
tunneling anisotropic spin polarization (TASP), which is defined
as\cite{Fabian2007:APS}
\begin{equation}\label{tasp-def}
    \textrm{TASP}_{[110]}(\phi)=\frac{P(0)-P(\phi)}{P(\phi)}.
\end{equation}

Taking into account that the Zeeman splitting in the normal metal
is small
%negligible in comparison to the exchange splitting in the
%ferromagnet
we can approximate $\kappa_{\sigma}\approx \kappa_{-\sigma}$. Then
for the DDM the conductance is given by Eq.~(\ref{cond-tot}). It
follows from Eqs.~(\ref{tamr-def}) and Eq.~(\ref{cond-tot}) that
the TAMR is given by
\begin{equation}\label{tamr-delta}
    \textrm{TAMR}_{[110]}(\phi)=\frac{G^{\textrm{aniso}}(0)-G^{\textrm{aniso}}(\phi)}{G^{\textrm{iso}}+
    G^{\textrm{aniso}}(\phi)}.
\end{equation}
For junctions in which the Bychkov-Rashba and Dresselhaus SOIs can
be considered as small perturbations, the anisotropy is small and
$G^{\textrm{aniso}}(\phi)\ll /G^{\textrm{iso}}$. In addition, the
spin-orbit contribution $G'=G_{\uparrow}'+G_{\downarrow}'$ [see
Eqs.~(\ref{g-app})-(\ref{g-0})]) to the isotropic part of the
conductance is also much smaller than the contribution
$G^{(0)}=G_{\uparrow}^{(0)}+G_{\downarrow}^{(0)}$, corresponding
to the system in the absence of the SOI, i.e., $G'\ll G^{(0)}$.
Therefore one can drop the contributions $G'$ and
$G^{\textrm{aniso}}$ from the denominator in
Eq.~(\ref{tamr-delta}). The TAMR can then be approximated as
\begin{equation}\label{tamr-d-approx}
    \textrm{TAMR}_{[110]}(\phi)\approx \frac{G^{\textrm{aniso}}(0)-G^{\textrm{aniso}}(\phi)}{G^{(0)}}.
\end{equation}

The substitution of Eq.~(\ref{gani}) into
Eq.~(\ref{tamr-d-approx}) leads to
\begin{eqnarray}\label{tamr-phi}
    &&\textrm{TAMR}_{[110]}(\phi)\approx \nonumber \\
    &&\frac{e^2}{h}\frac{\left\langle g_{2\uparrow}k_{\parallel}^{2}\right\rangle_{\uparrow}+
    \left\langle g_{2\downarrow}k_{\parallel}^{2}\right\rangle_{\downarrow}}
    {G^{(0)}}\lambda_{\alpha}\lambda_{\gamma}[\cos(2\phi)-1].
\end{eqnarray}
Here we have introduced the dimensionless SOC parameters
$\lambda_{\alpha}=2m_{0}\bar{\alpha}/\hbar^2$ and
$\lambda_{\gamma}=2m_{0}\bar{\gamma}/\hbar^2$. The functions
$g_{2\uparrow}$ and $g_{2\downarrow}$ are given by
Eq.~(\ref{g2-ani}).

The expression above gives the angular dependence of the TAMR and
is consistent with the angular dependence experimentally observed
in Fa/GaAs/Au tunnel junctions.\cite{Moser2007:PRL} It also
suggests that bias-induced changes of the sign of the
Bychkov-Rashba parameter $\bar{\alpha}$ result in the inversion
(change of sign) of the TAMR (such an inversion has been
experimentally observed.\cite{Moser2007:PRL}) Furthermore, one can
see from Eq.~(\ref{tamr-phi}) that the amplitude of the TAMR is
governed by the product
$\lambda_{\alpha}\lambda_{\gamma}\propto\bar{\alpha}\bar{\gamma}$
and the averages $\langle
g_{2\sigma}k_{\parallel}^{2}\rangle_{\sigma}$
($\sigma=\uparrow,\downarrow$). When $\bar{\alpha}\bar{\gamma}=0$,
the two-fold TAMR is suppressed (the suppression of the TAMR was
also observed in Ref.\onlinecite{Moser2007:PRL}), i.e., as long as
other anisotropic effects such as uniaxial strain are not present,
the Bychkov-Rashba (or Dresselhaus) SOI alone cannot explain the
experimentally observed $C_{2v}$ symmetry of the TAMR. The TAMR
vanishes also if the spin polarization of both electrodes becomes
sufficiently small. In such a case $k_{F,\uparrow}\approx
k_{F,\downarrow}$ and $g_{2\sigma}$ vanishes [see
Eq.~(\ref{g2-ani})], resulting in the suppression of the TAMR. On
the contrary, Eq.~(\ref{tamr-phi}) predicts an increase of the
TAMR amplitude for F/S/NM tunnel junctions whose constituents
exhibit large values of $\bar{\alpha}\bar{\gamma}$ as well as a
large spin polarization in the magnetic electrode.

\begin{figure}
\includegraphics[width=6cm]{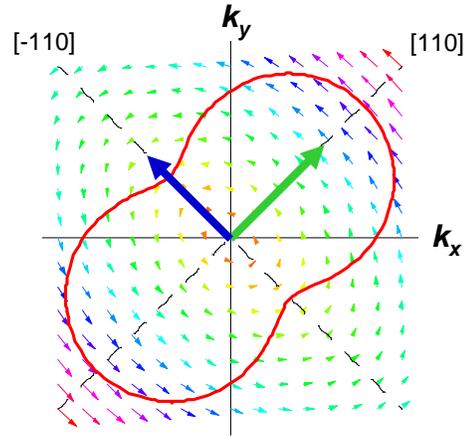}
\caption{(color online). (a) Schematics of the anisotropy of the
spin-orbit magnetic field $\mathbf{w}(\mathbf{k}_{\parallel})$.
Thin arrows represent a vector plot of the SOC field $\mathbf{w}$.
The solid line is a polar plot of the SOC field strength
$|\mathbf{w}(\mathbf{k}_{\parallel})|$ for a fixed value of
$k_{\parallel}=|\mathbf{k}_{\parallel}|$. When the magnetization
of the ferromagnet points along the [-110] direction [see thick
blue (dark) arrow], the direction of the strongest SOC field is
parallel to the incident majority spins which easily tunnel
through the barrier. On the contrary, for a magnetization
direction [110] [see thick green (light) arrow], the strongest SOC
field is perpendicular to the incident spins and the tunneling
becomes less favorable. The net result is a spin valve effect
whose efficiency depends upon the absolute orientation of the
magnetization and gives rise to the TAMR.} \label{fig-wfield}
\end{figure}

A simple, intuitive explanation of the origin of the uniaxial
anisotropy of the TAMR can be obtained by investigating the
dependence of the effective spin-orbit coupling field
$\mathbf{w}(\mathbf{k}_{\parallel})$ [see
Eq.~(\ref{rashba-delta})], i.e., the effective magnetic field that
the spins feel when traversing the semiconducting barrier. A
schematics of the anisotropy of the spin-orbit field
$\mathbf{w}(\mathbf{k}_{\parallel})$ is shown in
Fig.~(\ref{fig-wfield}), where the thin arrows represent a vector
plot of $\mathbf{w}(\mathbf{k}_{\parallel})$, while the solid line
is a polar plot of the field amplitude
$|\mathbf{w}(\mathbf{k}_{\parallel})|$ for a fixed value of
$k_{\parallel}=|\mathbf{k}_{\parallel}|$. The spin-orbit field is
oriented in the [110] ([-110]) direction at the points of low
(high) spin-orbit field, where the field amplitude $|\mathbf{w}|$
reaches a minimum (maximum). When the magnetization in the
ferromagnet points along the [-110] direction, the direction of
the highest spin-orbit field is parallel to the incident, majority
spins which are then easily transmitted through the barrier. On
the other hand, for a magnetization direction [110], the highest
spin-orbit field is perpendicular to the incident spins (to both
the majority and minority spins) and the transmission becomes less
favorable than in the case the magnetization is in the [-110]
direction. This spin-orbit induced difference in the tunneling
transmissivities depending on the magnetization direction results
in the uniaxial anisotropy of the TAMR.\cite{note2} Furthermore,
the magnetization direction dependence of the transmission and
reflection of the incident spins should be reflected in the local
density of states at the interfaces of the barrier. Within the DDM
the left (F/S) and right (S/NM) interfaces are merged into a
single plane and one can not distinguish between them. A more
detailed view of the role of the interfaces requires the use of
the SSOM. It turns out (this will be shown latter in this section)
that the F/S interface plays a major role in the TAMR phenomenon
while the S/NM interface appears irrelevant. This is intuitively
expected since the exchange splitting in the ferromagnet is much
larger than the Zeeman splitting in the normal metal.
Consequently, the spin-valve effect at the F/S interface is much
stronger than in the S/NM interface.

The local density of states reflect also the uniaxial anisotropy
of the TAMR with respect to the magnetization orientation in the
ferromagnet. In fact, one can introduce the anisotropic, local
density of states (ALDOS) through the definition
\begin{equation}\label{aldos}
    \textrm{ALDOS}_{[110]}(z,\phi)=\frac{\textrm{LDOS}(z,0)-\textrm{LDOS}(z,\phi)}{\textrm{LDOS}(z,\phi)},
\end{equation}
where
\begin{equation}\label{ldos}
    \textrm{LDOS}(z,\phi)=\sum_{\sigma=\uparrow,\downarrow}\int
    \frac{d\mathbf{k}_{\parallel}}{(2\pi)^2}\left|\Psi_{\sigma}(z,\phi,k_{\sigma
    F})\right|^{2},
\end{equation}
is the total local density of states at position $z$ and evaluated
at the Fermi surface determined by the Fermi wave vectors
\begin{equation}\label{kfermi}
    k_{\sigma F}=\sqrt{\frac{2m_{0}}{\hbar^2}\left(E_{F}+\sigma
    \frac{\Delta_{l}}{2}\right)-k_{\parallel}^{2}}.
\end{equation}
Since we are interested only in propagating states, we may
restrict the possible values of $k_{\parallel}$ to the interval
$[0,k_{max}^{\sigma}]$, with $k_{max}^{\sigma}$ given by
Eq.~(\ref{kmax}). Since the spin splitting in the normal metal
region is negligibly small,
$\kappa_{\sigma}\approx\kappa_{-\sigma}$. It follows from
Eqs.~(\ref{scattR}) and (\ref{trans}) that
\begin{equation}\label{trans-app}
    T_{\sigma}(E,k_{\parallel})\propto |\Psi_{\sigma}^{(r)}|^2,
\end{equation}
and, therefore, the conductance [see Eq.~(\ref{conductance})] is
related to the LDOS at $z=d$ as $G_{\sigma}(\phi)\propto
\textrm{LDOS}(d,\phi)$. One then obtains that
\begin{equation}\label{tamr-aldos}
    \textrm{TAMR}_{[110]}(\phi)\approx
    \textrm{ALDOS}_{[110]}(z=d,\phi).
\end{equation}

For a numerical illustration we consider an epitaxial Fe/GaAs/Au
heterojunction similar to that used in the experimental
observations reported in Ref.~\onlinecite{Moser2007:PRL}. We use
the value $m_{c}=0.067\;m_{0}$ for the electron effective mass in
the central (GaAs) region. The barrier width and hight (measured
from the Fermi energy) are, respectively, $d=80\;\textrm{\AA}$ and
$V_{c}=0.75 \textrm{ eV}$, corresponding to the experimental
samples in Ref.~\onlinecite{Moser2007:PRL}. For the Fe layer a
Stoner model with the majority and minority spin channels having
Fermi momenta $k_{F\uparrow}=1.05 \times 10^8 \textrm{ cm}^{-1}$
and $k_{F\downarrow}=0.44 \times 10^8 \textrm{ cm}^{-1}$,
\cite{Wang2003:JPCM} respectively, is assumed. The Fermi momentum
in Au is taken as $\kappa_{F}=1.2 \times 10^8 \textrm{ cm}^{-1}$.
\cite{Ashcroft:1976} We consider the case of relatively weak
magnetic fields (specifically, $B=0.5\textrm{ T}$). At high
magnetic fields, say, several Tesla, our model is invalid as it
does not include cyclotron effects which become relevant when the
cyclotron radius approaches the barrier width.

\begin{figure}
\includegraphics[width=7cm]{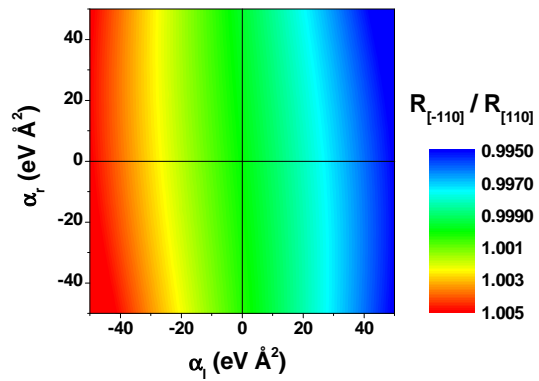}
\caption{(color online). Values of the ratio
$R_{[\bar{1}10]}/R_{[110]}$ as a function of the interface
Bychkov-Rashba parameters $\alpha_{l}$ and $\alpha_{r}$.}
\label{ag-ratio}
\end{figure}

The Dresselhaus spin-orbit parameter in GaAs is $\gamma \approx 24
\textrm{ eV
\AA}^{3}$.\cite{Winkler:2003,Ganichev2004:PRL,Fabian2007:APS} On
the other hand, the values of the interface Bychkov-Rashba
parameters $\alpha_{l}$, $\alpha_{r}$ [see
Eq.~(\ref{rashba-inter})] are not know for metal-semiconductor
interfaces. Due to the complexity of the problem, a theoretical
estimation of such parameters requires first-principles
calculations including the band structure details of the involved
materials,\cite{Gmitra2008} which is beyond the scope of the
present paper. Here we assume $\alpha_{l}$ and $\alpha_{r}$ as
phenomenological parameters which must be understood as the values
of the interface Bychkov-Rashba parameters at the
ferromagnet/semiconductor and semiconductor/normal metal
interfaces, respectively, averaged over all the relevant bands
contributing to the transport across the corresponding interfaces.
In order to investigate how does the degree of anisotropy depend
on these two parameters we performed calculations of the ratio
$R_{[\bar{1}10]}/R_{[110]}$ (which is a measure of the degree of
anisotropy\cite{Moser2007:PRL}) as a function of $\alpha_{l}$ and
$\alpha_{r}$ by using the spin-orbit Slonczewski model described
in Sec. \ref{ssom}. The results are shown in Fig.~\ref{ag-ratio},
where one can appreciate that the size of this ratio (and,
consequently, of the TAMR) is dominated by $\alpha_{l}$. This is
because the Zeeman splitting in Au is very small compared to the
exchange splitting in Fe and, consequently, the spin flips mainly
when crossing the ferromagnet/semiconductor interface. Then, since
the values of the TAMR are not very sensitive to the changes of
$\alpha_{r}$ we can set this parameter, without loss of
generality, to zero. This leaves $\alpha_{l}$ as a single fitting
parameter when comparing to experiment. The values of the
phenomenological parameter $\alpha_{l}$ were determined in
Refs.~\onlinecite{Moser2007:PRL,Fabian2007:APS} by fitting the
theory to the experimental value of the ratio
$R_{[-110]}/R_{[110]}$ and a very satisfactory agreement between
theory and experiment was achieved. This fitting at a single angle
was enough for the theoretical model to reproduce the {\it
complete} angular dependence of the TAMR, demonstrating the
robustness of the model.

\begin{figure}
\includegraphics[width=6cm]{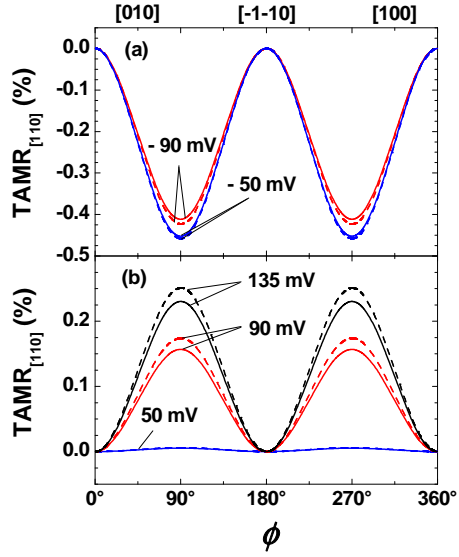}
\caption{(color online). Angular dependence of the TAMR in a
Fe/GaAs/Au MTJ. Solid and dashed curves represent the results
obtained within the SSOM and the DDM, respectively. (a) for
voltages -90 and -50 mV. (b) for voltages 50, 90, and 135 mV.}
\label{fig-tamrfsn}
\end{figure}

Assuming that the interface Bychkov-Rashba parameter $\alpha_{l}$
is voltage dependent and performing the fitting procedure for
different values of the bias voltage the bias dependence of
$\alpha_{l}$ can be extracted.\cite{Moser2007:PRL,Fabian2007:APS}
Here we use the same values of $\alpha_{l}$ reported in
Refs.~\onlinecite{Moser2007:PRL,Fabian2007:APS} for computing the
angular dependence of the TAMR at different values of the bias
voltage. The results are shown in Fig.~\ref{fig-tamrfsn}, where
the dashed and solid lines correspond to calculations within the
DDM and the SSOM, respectively. An overall agreement between the
two models can be appreciated. The TAMR exhibits an oscillatory
behavior as a function of the magnetization direction [see also
Eq.~(\ref{tamr-phi})] and can be inverted by changing the bias
voltage [compare Figs.~\ref{fig-tamrfsn}(a) and (b)]. This bias
induced inversion of the TAMR was experimentally observed in
Fe/GaAs/Au tunnel junctions\cite{Moser2007:PRL} and was explained
to occur as a consequence of a bias induced change in the sign of
the effective Bychkov-Rashba parameter. Preliminary {\it ab
initio} calculations\cite{Gmitra2008} for Fe/GaAs structures
suggest that the Bychkov-Rashba parameters associated with
different bands may have different values and even change the
sign. Thus, the effective value of the interface Bychkov-Rashba
parameter $\alpha_{l}$ will depend on which bands are the ones
that mainly contribute to the transport across the Fe/GaAs
interface (at low temperature those are the ones which have the
appropriate symmetry and lie inside the voltage window around the
Fermi energy). For different values of the bias voltage different
set of bands will be relevant to transport. On the other hand, to
different set of bands correspond different effective values of
the interface Bychkov-Rashba parameter. Consequently, $\alpha_{l}$
becomes strongly dependent on the bias voltage. The above analysis
leads to the conclusion that the inversion of the TAMR originates
from the bias induced sign change of the effective Bychkov-Rashba
SOI at the Fe/GaAs interface, as proposed in
Refs.~\onlinecite{Moser2007:PRL,Fabian2007:APS}.

\begin{figure}
\includegraphics[width=6cm]{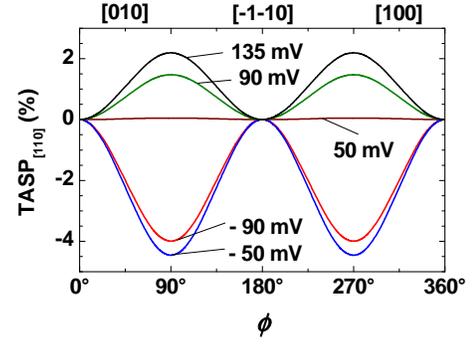}
\caption{(color online). Angular dependence of the TASP in a
Fe/GaAs/Au MTJ for different values of the bias voltage.}
\label{fig-aldos}
\end{figure}

The sign change of the effective Bychkov-Rashba parameter has also
influence on the tunneling spin polarization, resulting in the
bias induced inversion (change of sign) of the TASP as shown in
Fig.~\ref{fig-aldos}. Bias induced changes of the sign of the
tunneling spin polarization in Fe/GaAs/Cu MTJs has also been
reported.\cite{Chantis2007:PRL} The anisotropy of the tunneling
spin polarization, which also exhibits a two-fold symmetry,
indicates that the amount of transmitted and reflected spin at the
interfaces depends on the magnetization direction in the Fe layer,
resulting in an anisotropic local density of states at the Fermi
surface.
% [see Fig.~\ref{fig-aldos}(b)].
This is consistent with previous
works\cite{Gould2004:PRL,Ruster2005:PRL,Park2008:PRL} in which the
anisotropy of the density of states with respect to the
magnetization direction was related to the origin of the TAMR. In
fact, our model calculations reveal that the TAMR, ALDOS [see
Eq.~(\ref{tamr-aldos})], and TASP all exhibit a $C_{2v}$ symmetry
with the same kind of angular dependence (compare
Figs.~\ref{fig-tamrfsn} and \ref{fig-aldos}).

\begin{figure}
\includegraphics[width=6cm]{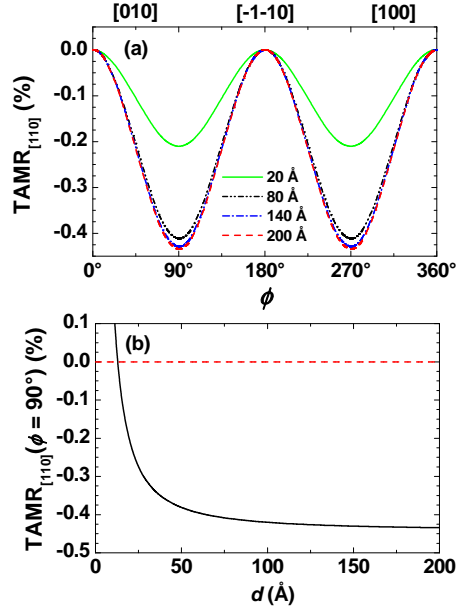}
\caption{(color online). (a) Angular dependence of the TAMR in a
Fe/GaAs/Au MTJ for different values of the barrier width $d$ and a
bias voltage $V_{_{bias}}=-90$ mV. (b) Amplitude of the TAMR (at
$\phi = 90^{\circ}$) as a function of the barrier width. The TAMR
curves were obtained by using the SSOM.} \label{fig-ddep}
\end{figure}

A system parameter that can influence the size of the TAMR is the
width of the barrier. The angular dependence of the TAMR
calculated within the SSOM for the case of $V_{bias}=-90\textrm{
meV}$ is displayed in Fig.~\ref{fig-ddep}(a) for different values
of the barrier width $d$. As one would expect, the changes in the
barrier width do not affect the two-fold symmetry of the TAMR but
only its amplitude, whose absolute value is predicted to increase
when increasing the width of the barrier [see
Fig.~\ref{fig-ddep}(b)].

We remark that our model neglects the contribution of the
spin-orbit-induced symmetries of the involved bulk structures.
Say, Fe exhibits a four-fold anisotropy, which should be reflected
in the tunneling density of states. The fact that this is not seen
in the experiment suggests that this effect is smaller than the
two-fold symmetry considered in our model.

\subsection{TAMR in ferromagnet/semiconductor/ferromagnet tunnel
junctions} \label{tamr-fsf}

\begin{figure}
\includegraphics[width=7cm]{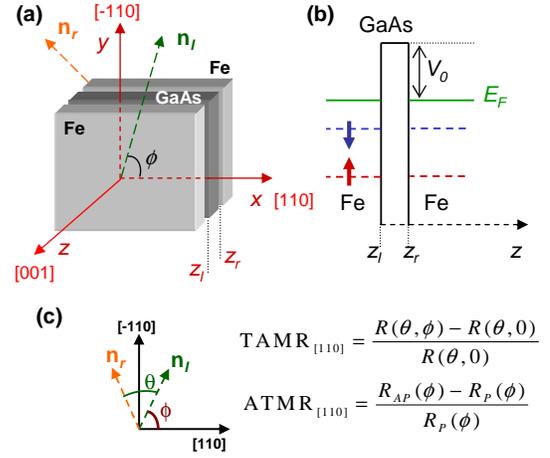}
\caption{(color online). (a) Schematics of a Fe/GaAs/Fe MTJ. The
magnetization direction in the left (right) ferromagnet is
specified by the vector $\mathbf{n}_{l}$ ($\mathbf{n}_{r}$). (b)
Schematics of the potential profile of the heterojunction along
the [001] direction. (c) Definition of the TAMR and ATMR in F/S/F
junctions. Here $R_{_P}(\phi)=R(0,\phi)$ and
$R_{_{AP}}(\phi)=R(180^{\circ},\phi)$.} \label{fig-fsf}
\end{figure}

Our discussion in Sec.~\ref{tamr-fsn} suggests that in the case of
a F/S/F tunnel junction, the magnetoresistance will depend on the
absolute direction of the magnetization in each of the
ferromagnets. In the left and right electrodes, the magnetization
directions with respect to the [110] crystallographic direction
are given by the angles $\phi=\phi_{l}$ and $\phi_{r}$,
respectively. For convenience we introduce the angle
$\theta=\phi_{r}-\phi_{l}$, describing the magnetization direction
in the right ferromagnet relative to that in the left ferromagnet
(see Fig.~\ref{fig-fsf}). Different values of the tunneling
magnetoresistance are expected to occur when in-plane rotating the
magnetizations of both ferromagnets at the same time, while
keeping the relative angle $\theta$ fixed. Thus, the expression
for the TAMR in F/S/NM junctions [see Eq.~(\ref{tamr-def})] can
now be generalized as
\begin{equation}\label{tamr-def-fsf}
    \textrm{TAMR}_{[110]}(\theta,\phi)=\frac{R(\theta,\phi)-R(\theta,0)}{R(\theta,0)}=
    \frac{G(\theta,0)-G(\theta,\phi)}{G(\theta,\phi)}.
\end{equation}

Following the same procedure as in Sec.~\ref{tamr-fsn}, one can,
in principle, obtain analytical expressions for the TAMR in a
F/S/F junction. It turns out however that in the general case
defined by Eq.~(\ref{tamr-def-fsf}) the resulting relations are
quite lengthy and not much simpler than the more accurate
expressions obtained within the SSOM (see Appendix \ref{appB}).
Therefore, we omit here the expressions resulting from the DDM and
show only the results obtained within the SSOM.

\begin{figure}
\includegraphics[width=7cm]{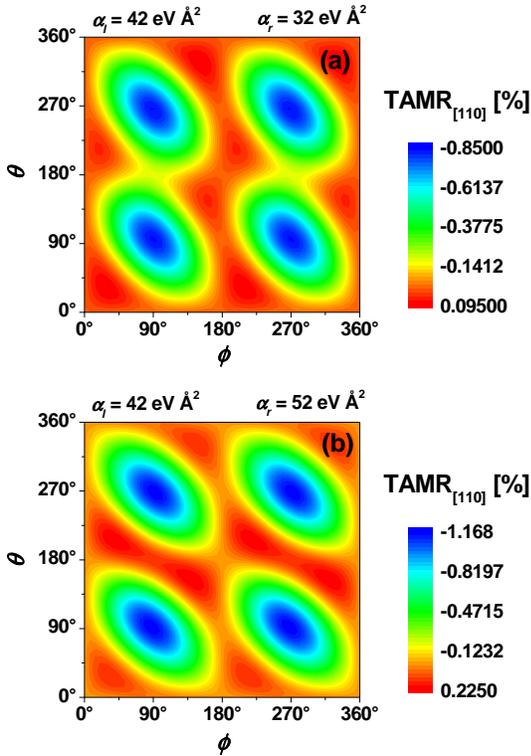}
\caption{(color online). Calculated TAMR defined in
Eq.~(\ref{tamr-def-fsf}) in a Fe/GaAs/Fe MTJ as a function of the
angles $\phi$ and $\theta$. (a) for $\alpha_{l}=42$ eV \AA$^2$ and
$\alpha_{r}=32$ eV \AA$^2$ (i.e., $\bar{\alpha}>0$). (b)for
$\alpha_{l}=42$ eV \AA$^2$ and $\alpha_{r}=52$ eV \AA$^2$ (i.e.,
$\bar{\alpha}<0$).} \label{fig-tamrg}
\end{figure}

\begin{figure}
\includegraphics[width=8cm]{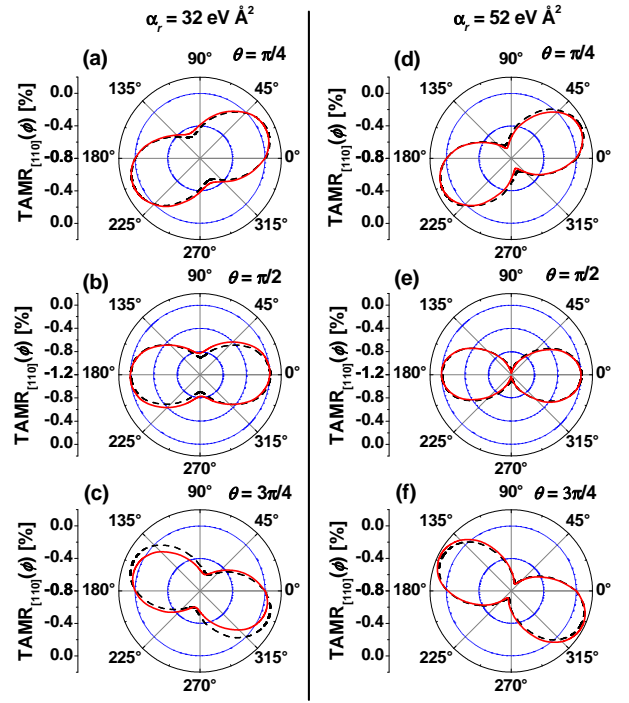}
\caption{(color online). Calculated angular dependence of the TAMR
in a Fe/GaAs/Fe MTJ for fixed values of the angle $\theta$ between
the electrode and counterelectrode magnetizations. The figures
correspond to polar plots of the traces $\theta=\pi/4$,
$\theta=\pi/2$ and $\theta=3\pi/4$ of Fig.~(\ref{fig-tamrg}).
Solid lines correspond to the calculations within the SSOM while
dashed lines are obtained from the phenomenological model by using
Eqs.~(\ref{tamr-gen-fsf}) - (\ref{b-def}).} \label{fig-trace}
\end{figure}

The dependence of the TAMR on the angles $\theta$ and $\phi$ is
shown in Fig.~\ref{fig-tamrg} for the case of a Fe/GaAs/Fe MTJ.
The width of the barrier is $d=80$ \AA. Two different cases,
corresponding to $\bar{\alpha}>0$ (a) and $\bar{\alpha}<0$ (b),
are considered. In both cases, the absolute value of the TAMR
reaches its maximal amplitude when the magnetization of the left
electrode is parallel to the [-110] direction (i.e.,
$\phi=90^{\circ},270^{\circ}$) and the one of the right electrode
is perpendicular to it (i.e., $\theta=90^{\circ},270^{\circ}$).
This is because this configuration, in our parabolic model,
corresponds to the case of the stronger structure inversion
asymmetry and, consequently, to the largest absolute value of the
effective Bychkov-Rashba parameter $\bar{\alpha}$. We also note
that at a fixed value of $\theta$ the TAMR has a two-fold symmetry
with respect to $\phi$ and vice versa. It is clear, however, that
the angles $\theta$ and $\phi$ play different roles in the
symmetry of the TAMR, which manifest in the lack of mirror
symmetry with respect to the axis $\theta = \phi$. We have
investigated some traces of the TAMR displayed in
Fig.~\ref{fig-tamrg}. The results are shown in
Fig.~\ref{fig-trace} where we present polar plots of the TAMR as a
function of $\phi$ for fixed values of $\theta$. The solid lines
correspond to the calculations within the SSOM. The meaning of the
dashed lines will be explained in Sec.~\ref{pheno}. The
orientation of the symmetry axis of the two-fold symmetric TAMR is
determined by the relative angle $\theta$ rather than by the
relative values of the interface Bychkov-Rashba parameters (note
for the same $\theta$, the symmetry axis does not change its
orientation with $\alpha_{r}$). The amplitude of the TAMR is
bigger for the case of $\alpha_{r}=52$ eV \AA$^2$ than for
$\alpha_{r}=32$ eV \AA$^2$ and shows a strong dependence on
$\theta$.

\section{Anisotropic tunneling magnetoresistance (ATMR)}
\label{atmr-sec}

We now consider the SOI effects on the TMR for the case of a F/S/F
tunnel junction. The {\it conventional} TMR effect in
ferromagnet/insulator/ferromagnet (F/I/F) tunnel junctions relies
on the dependence of the magnetoresistance across the junction on
the relative magnetization directions in the different
ferromagnetic layers and their spin
polarizations\cite{Slonczewski1989:PRB,Fabian2007:APS} and is
usually defined as
\begin{equation}\label{tmr-def}
    \textrm{TMR}=\frac{R_{AP}-R_{P}}{R_{P}},
\end{equation}
where $R_{P}$ ($R_{AP}$) is the magnetoresistance measured when
the magnetization of the left and right ferromagnetic layers are
parallel (antiparallel).

In the case of F/S/F heterojunctions, the interference between
Bychkov-Rashba and Dresselhaus SOIs leads to anisotropic effects.
Consequently, the conventional TMR in F/S/F junctions will depend
not only on the relative but also on the absolute magnetization
directions in the ferromagnets, resulting in an anisotropic
magnetoresistance. For quantifying this phenomenon we use the
following anisotropic generalization of the TMR,
\begin{equation}\label{atmr-def}
    \textrm{ATMR}_{[110]}=\frac{R_{AP}(\phi)-R_{P}(\phi)}{R_{P}(\phi)}=
    \frac{G_{P}(\phi)-G_{AP}(\phi)}{G_{AP}(\phi)},
\end{equation}
which now accounts for the magnetoresistance dependence on the
absolute magnetization orientations with respect to the [110]
crystallographic direction (see Fig.~\ref{fig-fsf}). Furthermore,
the efficiency $\eta$ of the anisotropic effects on the tunneling
magnetoresistance can be defined as
\begin{equation}\label{effi}
    \eta=\frac{\textrm{ATMR}_{[110]}(\phi)-\textrm{ATMR}_{[110]}(0)}{\textrm{ATMR}_{[110]}(0)}.
\end{equation}

A simplified approximate expression for the ATMR can be found
within the DDM by following the same procedure as in
Sec.~\ref{tamr-fsn}. The result is
\begin{eqnarray}\label{atmr-res}
    &&\textrm{ATMR}_{[110]}\approx \nonumber \\
    &&\textrm{TMR}-\frac{e^2}{h G_{AP}^{(0)}}\left[
    \frac{\left(\lambda_{\alpha}^{2}+\lambda_{\gamma}^{2}\right)}
    {2}
    \sum_{\sigma=\uparrow,\downarrow}\left\langle
    (g_{1\sigma}^{P}-g_{1\sigma}^{AP}
    )k_{\parallel}^{2}\right\rangle_{\sigma}\right. \nonumber \\
    &&+\left.\lambda_{\alpha}\lambda_{\gamma}\cos(2\phi)
    \sum_{\sigma=\uparrow,\downarrow}\left\langle
    (g_{2\sigma}^{AP}-g_{2\sigma}^{P})k_{\parallel}^{2}\right\rangle_{\sigma}\right],
\end{eqnarray}
where $\textrm{TMR}=(G_{P}^{(0)}-G_{AP}^{(0)})/G_{AP}^{(0)}$ is
the conventional TMR in the absence of SOI, and the functions
$g_{1\sigma}^{P}$, $g_{2\sigma}^{P}$, $g_{1\sigma}^{AP}$, and
$g_{2\sigma}^{AP}$ are given by the
Eqs.~(\ref{g1-iso-p}),(\ref{g2-ani-p}), (\ref{g1-iso-ap}), and
(\ref{g2-ani-ap}), respectively. Assuming that the spin-orbit
effects on the conventional TMR are small one can approximate the
ATMR efficiency as
\begin{equation}\label{eta-ddm}
    \eta \approx \frac{e^2}{h G_{AP}^{(0)}}
    \frac{\lambda_{\alpha}\lambda_{\gamma}[\cos(2\phi)-1]}{\textrm{TMR}}
    \sum_{\sigma=\uparrow,\downarrow}\left\langle
    (g_{2\sigma}^{P}-g_{2\sigma}^{AP})k_{\parallel}^{2}\right\rangle_{\sigma}.
\end{equation}
Note that the angular dependence of the efficiency $\eta$ is
similar to that of the TAMR given in Eq.~(\ref{tamr-phi}).

\begin{figure}
\includegraphics[width=5cm]{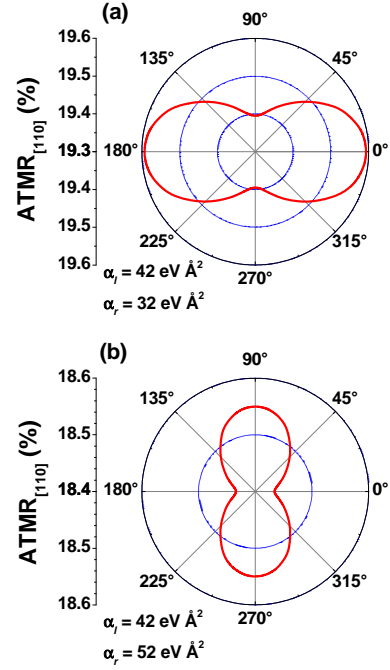}
\caption{(color online). Angular dependence of the ATMR defined in
Eq.~(\ref{atmr-def}) for a Fe/GaAs/Fe MTJ. (a) for $\alpha_{l}=42$
eV \AA$^2$ and $\alpha_{r}=32$ eV \AA$^2$ (i.e.,
$\bar{\alpha}>0$). (b)for $\alpha_{l}=42$ eV \AA$^2$ and
$\alpha_{r}=52$ eV \AA$^2$ (i.e., $\bar{\alpha}<0$).}
\label{fig-atmr}
\end{figure}

It follows from Eq.~(\ref{atmr-res}) that when both Bychkov-Rashba
and Dresselhaus SOIs are present (i.e., when
$\lambda_{\alpha}\lambda_{\gamma}\propto\bar{\alpha}\bar{\gamma}\neq
0$), the magnetoresistance becomes anisotropic, resulting in the
ATMR effect here predicted. One can see also that this effect
exhibits a two-fold symmetry with a $\cos(2\phi)$ angular
dependence. Unlike the TAMR, when $\bar{\alpha}\bar{\gamma}=0$ the
ATMR becomes isotropic but does not vanish. However, in such a
limit the efficiency $\eta$ of the ATMR do vanish [see
Eq.~(\ref{eta-ddm})], indicating the absence of anisotropy. Like
for the TAMR, changes of the sign of $\bar{\alpha}$ result in the
inversion (change of sign) of the efficiency $\eta$ of the ATMR.
On the other hand, since the TMR contribution in
Eq.~(\ref{atmr-res}) is usually the dominant one, the bias-induced
changes of the sign of the Bychkov-Rashba parameter $\bar{\alpha}$
will not, in general, cause the inversion (change of sign) of the
ATMR. However, the fact that the TMR contribution may change sign
in dependence of the applied
voltage\cite{DeTeresa1999:PRL,Sharma1999:PRL,Moser2006:APL} can
result in the inversion of the ATMR.

We performed calculations within the SSOM of the angular
dependence of the ATMR for a Fe/GaAs/Fe MTJ with a fixed value of
the left-interface Bychkov-Rashba parameter ($\alpha_{l}=42$ eV
\AA$^2$) corresponding to a bias voltage $V_{_{bias}}=-90$ mV .
The results are shown in Figs.~\ref{fig-atmr} (a) and (b) for the
parameters $\alpha_{r}=32$ eV \AA$^2$ and $\alpha_{r}=52$ eV
\AA$^2$. The angular dependence of the ATMR is consistent with
Eq.~(\ref{atmr-res}). The orientation of the symmetry axis of the
ATMR is determined by the sign of the effective value
$\bar{\alpha}$. On the other hand, the amplitude of the ATMR shows
a behavior opposite to the one corresponding to the TAMR (see
Fig.~\ref{fig-trace}), in the sense that now the amplitude of the
anisotropy of the TMR is bigger for the case $\alpha_{r}=32$ eV
\AA$^2$ than for $\alpha_{r}=52$ eV \AA$^2$ and, unlike for the
case of the TAMR (see Fig.~\ref{fig-trace}), the orientation of
the symmetry axis of the ATMR can be flipped by varying the value
of $\alpha_{r}$.

\section{Phenomenological model of TAMR and ATMR}
\label{pheno}

In order to explain the origin of the angular dependence of the
TAMR [see Eq.~(\ref{tamr-phi})] in F/S/NM junctions, a
phenomenological model based on rather general symmetry
considerations was proposed in Ref.\onlinecite{Moser2007:PRL} and
elaborated in more details in Ref.\onlinecite{Fabian2007:APS}.
Here we extend this phenomenological model to the case of F/S/F
tunnel junctions.

For a given $\mathbf{k}_{\parallel}$, there are three preferential
directions in the system: i) the magnetization direction
$\mathbf{n}_{l}$ in the left ferromagnet, ii) the magnetization
direction $\mathbf{n}_{r}$ in the right ferromagnet, and iii) the
direction of the effective spin-orbit field $\mathbf{w}$. A scalar
quantity such as the transmissivity for the $\sigma$ spin channel
can be expanded in a series of the all possible scalars one can
form with the vectors $\mathbf{n}_{l}$, $\mathbf{n}_{r}$, and
$\mathbf{w}$. Thus, to the second order in $|\mathbf{w}|$, we have
\begin{equation}\label{t-expa}
    T_{\sigma}=T_{\sigma}^{(0)}+T_{\sigma}^{(2)}.
\end{equation}
Note that because of symmetry reasons,\cite{Fabian2007:APS} the
linear in $|\mathbf{w}|$ terms vanishes after averaging over
$k_{\parallel}$ and, therefore, we have omitted them in
Eq.~(\ref{t-expa}). The zero order terms have the general form
\begin{equation}\label{t-0}
    T_{\sigma}^{(0)}=a_{1\sigma}^{(0)}
    %(\mathbf{n}_{l}\cdot\mathbf{n}_{l})+
    %a_{2\sigma}^{(0)}(\mathbf{n}_{r}\cdot\mathbf{n}_{r})
    +a_{2\sigma}^{(0)}(\mathbf{n}_{l}\cdot\mathbf{n}_{r}),
\end{equation}
where $a_{1\sigma}^{(0)}$
%, $a_{2\sigma}^{(0)}$,
and $a_{2\sigma}^{(0)}$ are isotropic expansion coefficients.

Taking into account that
$\mathbf{n}_{l}=[\cos(\phi+\pi/4),\sin(\phi+\pi/4),0]$ and
$\mathbf{n}_{r}=[\cos(\theta+\phi+\pi/4),\sin(\theta+\phi+\pi/4),0]$,
the general angular dependence of $T^{(0)}$ can be extracted from
Eq.~(\ref{t-0}). The result is
\begin{equation}\label{t-0-f}
    T_{\sigma}^{(0)}=a_{1\sigma}^{(0)}
    %+a_{2\sigma}^{(0)})
    +a_{2\sigma}^{(0)}\cos(\theta).
\end{equation}
This equation describes the dependence of the transmissivity on
the relative angle $\theta$ between the magnetizations in the left
and right ferromagnetic electrodes in the absence of SOI and is
consistent with previous
results.\cite{Slonczewski1989:PRB,Fabian2007:APS}

The second order contribution $T^{(2)}$ can be cast in the
following general form
\begin{eqnarray}\label{t-2}
    T_{\sigma}^{(2)}&=&
    [a_{1\sigma}^{(2)}+a_{2\sigma}^{(2)}(\mathbf{n}_{l}\cdot\mathbf{n}_{r})]|\mathbf{w}|^{2}
    + a_{3\sigma}^{(2)}|(\mathbf{n}_{l}\cdot\mathbf{w})|^{2} \nonumber \\
    &+&a_{4\sigma}^{(2)}|(\mathbf{n}_{r}\cdot\mathbf{w})|^{2}+
    a_{5\sigma}^{(2)}(\mathbf{n}_{l}\cdot\mathbf{w})(\mathbf{n}_{r}\cdot\mathbf{w}),
\end{eqnarray}
with the expansion coefficients $a_{i\sigma}^{(2)}$
($i=1,2,...,5$) being angular independent. To second order in the
SOI, the conductance is then given by
\begin{equation}\label{cond-feno}
    G_{\sigma}=\frac{e^2}{h}\left(\langle T_{\sigma}^{(0)} \rangle_{\mathbf{k}_{\parallel}}+\langle T_{\sigma}^{(2)}
    \rangle_{\mathbf{k}_{\parallel}}\right),
\end{equation}
where $\langle ...\rangle_{\mathbf{k}_{\parallel}}$ denotes
average over $\mathbf{k}_{\parallel}$ and evaluation at the Fermi
energy $E_{F}$.

Taking into account Eqs.~(\ref{rashba-delta}), (\ref{t-0-f}),
(\ref{t-2}), and (\ref{cond-feno}) and performing the
corresponding angular averaging in $\mathbf{k}$-space one obtains
the following, general form of the total conductance
$G=G_{\uparrow}+G_{\downarrow}$,
\begin{eqnarray}\label{cond-gen}
    G(\theta,\phi)&=&G_{1}+G_{2}\cos(\theta)+G_{3}\cos(2\phi)\nonumber
    \\
    &+&G_{4}\cos(2\phi+2\theta)+ G_{5}\cos(2\phi+\theta),
\end{eqnarray}
where
\begin{eqnarray}\label{g1g-feno}
    &&G_{1}= \frac{e^2}{h}\times \nonumber \\
    &&\sum_{\sigma=\uparrow,\downarrow}\left\langle a_{1\sigma}^{(0)}+
    %a_{2\sigma}^{(0)}+
    \frac{|\mathbf{w}|^{2}}{2}
    (2a_{1\sigma}^{(2)}+
    a_{3\sigma}^{(2)}+a_{4\sigma}^{(2)})k_{\parallel}^{2}\right\rangle_{\sigma},
\end{eqnarray}
\begin{equation}\label{g2g-feno}
    G_{2}=\frac{e^2}{h}\sum_{\sigma=\uparrow,\downarrow}\left\langle a_{2\sigma}^{(0)}
    +\frac{|\mathbf{w}|^{2}}{2}(2a_{2\sigma}^{(2)}+
    a_{5\sigma}^{(2)})k_{\parallel}^{2}\right\rangle_{\sigma},
\end{equation}
\begin{equation}\label{g3g-feno}
    G_{3}=\frac{e^2}{h}\sum_{\sigma=\uparrow,\downarrow}
    \left\langle a_{3\sigma}^{(2)}w_{x}w_{y}k_{\parallel}^{2}\right\rangle_{\sigma},
\end{equation}
\begin{equation}\label{g4g-feno}
    G_{4}=\frac{e^2}{h}\sum_{\sigma=\uparrow,\downarrow}
    \left\langle a_{4\sigma}^{(2)}w_{x}w_{y}k_{\parallel}^{2}\right\rangle_{\sigma},
\end{equation}
and
\begin{equation}\label{g5g-feno}
    G_{5}=\frac{e^2}{h}\sum_{\sigma=\uparrow,\downarrow}
    \left\langle a_{5\sigma}^{(2)}w_{x}w_{y}k_{\parallel}^{2}\right\rangle_{\sigma}.
\end{equation}
In the equations above the average $\langle ... \rangle_{\sigma}$
have the same meaning as in Eq.~(\ref{aver}). We note that for
systems which are isotropic in the absence of the SOI these
equations are quite general. They are valid up to second order in
the strength of the spin-orbit coupling field regardless of the
specific form of $\mathbf{w}$.

One can see that for the case of a F/S/NM junction or for the case
of parallel and antiparallel configurations in a F/S/F junction,
the general angular dependence given in Eq.~(\ref{cond-gen}) is
consistent with the corresponding expressions obtained in Appendix
\ref{appA}. We want to stress, however, that the relations
presented in Appendix \ref{appA} are the result of a specific
approximation (the DDM), while Eqs.~(\ref{cond-gen}) -
(\ref{g5g-feno}) are general, and valid for any model or
approximation (as far as the system without SOI is isotropic, the
effective spin-orbit field is weak, and $\mathbf{n}_{l}$,
$\mathbf{n}_{r}$, and $\mathbf{w}$ lie in a plane perpendicular to
the current flow). More general relations corresponding to
arbitrary orientations of $\mathbf{n}_{l}$, $\mathbf{n}_{r}$, and
$\mathbf{w}$ will be given elsewhere.\cite{Matos2008}

The general form of the TAMR in F/S/NM junctions follows from
Eqs.~(\ref{rashba-delta}), (\ref{tamr-def}) and (\ref{cond-gen}).
Proceeding as in Sec.~\ref{tamr-fsn} one finds
\begin{eqnarray}\label{tamr-gen}
    &&\textrm{TAMR}_{[110]}(\phi)\approx \nonumber \\
    &&\frac{e^2}{h}\frac{1}{G^{(0)}}
    \left[\sum_{\sigma=\uparrow,\downarrow}
    \left\langle -(a_{3\sigma}^{(2)}+a_{4\sigma}^{(2)}+a_{5\sigma}^{(2)})
    k_{\parallel}^{2}\right\rangle_{\sigma}\right]\times \nonumber \\
    &&\lambda_{\alpha}\lambda_{\gamma}[\cos(2\phi)-1],
\end{eqnarray}
which is consistent with the corresponding expression found within
the DDM [see Eq.~(\ref{tamr-phi})]. Similarly, one obtains for the
TMR and ATMR in F/S/F junctions the following general expressions,
\begin{equation}\label{tmr-gen}
    \textrm{TMR}=\frac{e^2}{h}\frac{1}{G_{AP}^{(0)}}\sum_{\sigma=\uparrow,\downarrow}
    \left\langle 2a_{3\sigma}^{(0)}
    k_{\parallel}^{2}\right\rangle_{\sigma},
\end{equation}
and
\begin{eqnarray}\label{atmr-gen}
    &&\textrm{ATMR}_{[110]}(\phi)= \nonumber \\
    && \textrm{TMR}+\frac{e^2}{2h}\frac{(\lambda_{\alpha}^{2}+\lambda_{\gamma}^{2})}{G_{AP}^{(0)}}
    \sum_{\sigma=\uparrow,\downarrow}
    \left\langle (4a_{2\sigma}^{(2)}+2a_{5\sigma}^{(2)})
    k_{\parallel}^{2}\right\rangle_{\sigma} \nonumber \\
    &&+
    \frac{e^2}{h}\frac{1}{G_{AP}^{(0)}}\left[\sum_{\sigma=\uparrow,\downarrow}
    \left\langle 2a_{4\sigma}^{(2)}
    k_{\parallel}^{2}\right\rangle_{\sigma}\right]\lambda_{\alpha}\lambda_{\gamma}\cos(2\phi),
\end{eqnarray}
respectively. Note that Eq.~(\ref{atmr-gen}) is consistent with
our previous result in Eq.~(\ref{atmr-res}).

The TAMR in a F/S/F exhibits a more complicated angular
dependence, which has the general form
\begin{eqnarray}\label{tamr-gen-fsf}
    &&\textrm{TAMR}_{[110]}(\theta,\phi)= \nonumber \\
    &&\frac{A(\theta)[1-\cos(2\phi)]+
    B(\theta)\sin(2\phi)}{G_{1}+G_{2}\cos\theta+A(\theta)\cos(2\phi)-B(\theta)\sin(2\phi)},
\end{eqnarray}
where
\begin{equation}\label{a-def}
    A(\theta)=G_{3}+G_{4}\cos(2\theta)+G_{5}\cos\theta,
\end{equation}
and
\begin{equation}\label{b-def}
    B(\theta)=G_{4}\sin(2\theta)+G_{5}\sin\theta.
\end{equation}

At the phenomenological level, the expansion constants
$G_{1},...G_{5}$ in Eqs.~(\ref{cond-gen}) and (\ref{tamr-gen-fsf})
can be determined from conductance measurements (or conductance
theoretical evaluation) at selected values of $\theta$ and $\phi$.
For example, the values of the phenomenological constants can be
determine as follows
\begin{equation}\label{g1-cond}
    G_{1}=\frac{G(\pi/2,0)+G(\pi/2,\pi/2)}{2},
\end{equation}
\begin{equation}\label{g2-cond}
    G_{2}=G(0,\pi/4)-G_{1},
\end{equation}
\begin{equation}\label{g5-cond}
    G_{5}=G_{1}-G(\pi/2,\pi/4),
\end{equation}
\begin{equation}\label{g3-cond}
    G_{3}=G(\pi/4,0)-G_{1}-\frac{\sqrt{2}}{2}(G_{2}+G_{5}),
\end{equation}
and
\begin{equation}\label{g4-cond}
    G_{4}=G_{1}+G_{3}-G(\pi/2,0).
\end{equation}
It is worth noting that although we have referred to the
coefficients $G_{1},...,G_{5}$ as constants, strictly speaking,
these parameters may exhibit a dependence on the relative angle
$\theta$. Such a dependence may originate from the fact that the
averages containing the components of the spin-orbit field [see
Eqs.~(\ref{g1g-feno}) - (\ref{g5g-feno})] are, in principle,
$\theta$-dependent. However, as shown below, this effect is weak
for the system here considered.

In order to check the validity of the general angular dependence
given in Eqs.~(\ref{cond-gen}) and (\ref{tamr-gen-fsf}) we have
computed the expansion coefficients from Eqs.~(\ref{g1-cond}) -
(\ref{g4-cond}) by using the SSOM. We then evaluate the TAMR by
using the phenomenological relation given in
Eq.~(\ref{tamr-gen-fsf}) and compare the results (see dashed lines
in Fig.~\ref{fig-trace}) with the full angular dependence obtained
from the SSOM (solid lines in Fig.~\ref{fig-trace}). The agreement
is satisfactory although some small discrepancies appear. We
attribute these discrepancies to the fact that when determining
the coefficients $G_{1},...,G_{5}$ from Eqs.~(\ref{g1-cond}) -
(\ref{g4-cond}) we have ignored, as discussed above, that these
coefficients are weakly $\theta$-dependent.

\section{Conclusions}
\label{conclu}

We have investigated the spin-orbit induced anisotropy of the
tunneling magnetoresistance of F/S/NM and F/S/F MTJs. By
performing model calculations we have shown that the two-fold
symmetry of the TAMR effect may arise from the interplay of
Bychkov-Rashba and Dresselhaus SOIs. The spin-orbit interference
effects in F/S/F MTJs lead to the anisotropy of the conventional
TMR. Thus, the ATMR depends on the absolute magnetization
directions in the ferromagnets. The magnetoresistance dependence
on the absolute orientation of the magnetization in the
ferromagnets is deduced from general symmetry considerations.

\begin{acknowledgements}
%\section*{Acknowledgements}
We are grateful to M. Gmitra for useful hints and discussions and
to D. Weiss and J. Moser for discussions on TAMR related
experiments. This work was supported by the Deutsche
Forschungsgemeinschaft (DFG) through the Sonderforschungsbereich
(SFB) 689.
\end{acknowledgements}

\appendix
\section{}
\label{appA}

The eigenfunctions of the Hamiltonian given by Eq.~(\ref{hamilt})
obey the following matching conditions
\begin{widetext}
\begin{equation}\label{match-01}
    \Psi_{\sigma}^{(l)}(0^{-})=\Psi_{\sigma}^{(r)}(0^{+});
    \;\;
    \frac{\hbar^2}{2m_{0}}\left.\frac{d
    \Psi_{\sigma}^{(l)}}{dz}\right|_{z=0^{-}}+(V_{0}d+\mathbf{w}\cdot
    \mbox{\boldmath$\sigma$})\Psi_{\sigma}^{(l)}(0^{-})=
    \frac{\hbar^2}{2m_{0}}\left.\frac{d
    \Psi_{\sigma}^{(r)}}{dz}\right|_{z=0^{+}},
\end{equation}
\end{widetext}
which, with the scattering states in Eqs.(\ref{scattL}) and
(\ref{scattR}), lead to a system of 4 linear equations for
determining the coefficients $r_{\sigma,\sigma}$,
$r_{\sigma,-\sigma}$, $t_{\sigma,\sigma}$, and
$t_{\sigma,-\sigma}$. The transmission coefficients are found to
be given by
\begin{eqnarray}\label{tss0}
    t_{\sigma,\sigma}=&-&\frac{4d^2\sqrt{k_{\sigma}}
    \left(k_{-\sigma}+\kappa_{-\sigma}+iQ\right)\left(1+e^{i(\phi_{r}-\phi_{l})}\right)}{\Omega}\nonumber
    \\
    &+&\frac{8id\sqrt{k_{\sigma}}\left(\mathbf{U}\cdot
    \mathbf{S}_{\sigma,\sigma}\right)}{\Omega},
\end{eqnarray}
and
\begin{eqnarray}\label{tsf0}
    t_{\sigma,-\sigma}&=&\frac{4d\sqrt{k_{\sigma}d}
    \left(k_{-\sigma}+\kappa_{\sigma}+iQ\right)\left(1-e^{i(\phi_{r}-\phi_{l})}\right)}{\Omega}\nonumber
    \\
    &-&\frac{8i\left(\mathbf{U}\cdot
    \mathbf{S}_{\sigma,-\sigma}\right)}{\Omega},
\end{eqnarray}
where
\begin{equation}\label{omega}
    \Omega=\Omega_{+}(-)\Omega_{-}(+)-\Omega_{+}(+)\Omega_{-}(-),
\end{equation}
with
\begin{eqnarray}\label{omegas-pm}
    \Omega_{\pm}(\lambda)&=&d\left(k_{\pm\sigma}+\kappa_{\lambda\sigma}+iQ\right)
    \left(1\pm \lambda e^{i(\phi_{r}-\phi_{l})}\right)\nonumber \\
    &+&2i\left(\mathbf{U}\cdot
    \mathbf{S}_{\pm \sigma,\lambda\sigma}\right).
\end{eqnarray}
The vectors $\mathbf{S}_{\sigma,\sigma'}$ ($\sigma'=\pm \sigma$)
are given by
\begin{equation}\label{eses}
    \mathbf{S}_{\sigma,\sigma'}=\chi_{\sigma}^{(l)\dag}\mbox{\boldmath$\sigma$}\chi_{\sigma'}^{(r)},
\end{equation}
while $Q=2m_{0}V_{0}d/\hbar^{2}$ and
$\mathbf{U}=(2m_{0}d/\hbar^{2})\mathbf{w}$.

The transmissivity for an incident particle with spin $\sigma$ is
given by Eq.~(\ref{trans}). For the case of a F/S/NM junction we
can approximate $\kappa=\kappa_{\sigma}\approx \kappa_{-\sigma}$
and the transmissivity reduces to
\begin{equation}\label{trans-app}
    T_{\sigma}=\textrm{Re}\left[\kappa
    \left(|t_{\sigma,\sigma}|^{2}+|t_{\sigma,-\sigma}|^{2}\right)\right].
\end{equation}
In order to obtain a simplified analytical expression for the TAMR
in F/S/NM junctions, we consider the case in which the effective
spin-orbit field is small. In such a case one can expand
Eq.~(\ref{trans-app}) in powers of $w=|\mathbf{w}|$ and obtain the
conductance from Eq.~(\ref{conductance}). The result, up to second
order in $w$, reads
\begin{equation}\label{g-app}
    G_{\sigma}\approx
    G_{\sigma}^{\textrm{iso}}+G_{\sigma}^{\textrm{aniso}}.
\end{equation}
The isotropic part of the conductance is given by
\begin{equation}\label{giso-tot}
    G_{\sigma}^{\textrm{iso}}=G_{\sigma}^{(0)}+G'_{\sigma},
\end{equation}
where
\begin{equation}\label{g-0}
    G_{\sigma}^{(0)}=\frac{e^2}{h}\left\langle \frac{4|k_{F,\sigma}|\kappa_{F}}
    {|A_{\sigma}|^2} \right\rangle_{\sigma},
\end{equation}
with
\begin{equation}\label{a-s}
    A_{\pm \sigma}=k_{F,\pm \sigma}+\kappa_{F}+iQ,
\end{equation}
is the conductance in absence of the SOI and
\begin{equation}\label{gprime}
    G'_{\sigma}=-\frac{e^2}{2h}(\lambda_{\alpha}^{2}+
    \lambda_{\gamma}^{2})\langle g_{1\sigma}k_{\parallel}^{2} \rangle_{\sigma},
\end{equation}
is the isotropic contribution induced by the SOI. Here we have
denoted $k_{F\sigma}=k_{\sigma}(E_{F},k_{\parallel})$,
$\kappa_{F}=\kappa(E_{F},k_{\parallel})$ and
\begin{equation}\label{g1-iso}
    g_{1\sigma}=\frac{8\kappa_{F}|k_{F,\sigma}|}{|A_{\sigma}|^{2}|A_{-\sigma}|^{2}}
    \left(\frac{2\textrm{Re}[A_{\sigma}A_{-\sigma}]}{|A_{\sigma}|^2}-1\right)-g_{2\sigma}.
\end{equation}
We have also introduced dimensionless SOC parameters as
\begin{equation}\label{dimless-soc}
    \lambda_{\alpha}=\frac{2m_{0}}{\hbar^2}\bar{\alpha},\;\;\lambda_{\gamma}=\frac{2m_{0}}{\hbar^2}\bar{\gamma}.
\end{equation}
The function $g_2$ is defined below in Eq.~(\ref{g2-ani}). The
average
\begin{equation}\label{aver}
    \langle
...\rangle_{\sigma}=\frac{1}{2\pi}\int_{0}^{k_{_{\textrm{max}}}^{\sigma}}...\;k_{\parallel}dk_{\parallel},
\end{equation}
where
\begin{equation}\label{kmax}
    k_{_{\textrm{max}}}^{\sigma}=\textrm{min}\left(\sqrt{\frac{2m_{0}}{\hbar^2}
    \left(E_{F}+\sigma\frac{\Delta_{l}}{2}\right)},\sqrt{\frac{2m_{0}}{\hbar^2}E_{F}}\right),
\end{equation}
denotes the maximum values of $k_{\parallel}$ for which we have
incident and transmitted propagating states. In the average
defined in Eq.~(\ref{aver}) the corresponding angular integration
over $\mathbf{k}_{\parallel}$ [see Eq.~(\ref{conductance})] has
already been performed.

The spin-orbit induced anisotropy of the conductance is determined
by the relation
\begin{equation}\label{gani}
    G_{\sigma}^{\textrm{aniso}}(\phi)=-\frac{e^2}{h}\lambda_{\alpha}\lambda_{\gamma}\langle g_{2\sigma}k_{\parallel}^{2}
    \rangle_{\sigma}
    \cos(2\phi),
\end{equation}
where
\begin{widetext}
\begin{equation}\label{g2-ani}
    g_{2\sigma}=\frac{4\kappa_{F}|k_{F,\sigma}|\left(|A_{-\sigma}|^{2}|A_{\sigma}-A_{-\sigma}|^{2}-4\textrm{Im}[A_{-\sigma}](|A_{-\sigma}|^{2}
    \textrm{Im}[A_{\sigma}]-|A_{\sigma}|^{2}\textrm{Im}[A_{-\sigma}])\right)}{|A_{\sigma}|^{4}|A_{-\sigma}|^{4}}.
\end{equation}
\end{widetext}
The total conductance can then be written as
\begin{equation}\label{cond-tot}
    G(\phi)\approx G^{\textrm{iso}}+G^{\textrm{aniso}}(\phi),
\end{equation}
with
$G^{\textrm{iso}}=G_{\uparrow}^{\textrm{iso}}+G_{\downarrow}^{\textrm{iso}}$
and
$G^{\textrm{aniso}}(\phi)=G_{\uparrow}^{\textrm{aniso}}(\phi)+G_{\downarrow}^{\textrm{aniso}}(\phi)$.

We now consider the case of a F/S/F tunnel junction. For the
general case of arbitrary orientations of the magnetization in the
left and right ferromagnets, the analytical expressions for the
conductance are very lengthy and we therefore omit them here.
Simpler expressions are found, however, for the particular cases
of parallel (P) and antiparallel (AP) configurations. If we
consider that the ferromagnetic electrodes are made of the same
material, then $k_{\sigma}=\kappa_{\sigma}$ and the transmissivity
reduces to
\begin{equation}\label{trans-fsf}
    T_{\sigma}=\textrm{Re}\left[k_{\sigma}
    |t_{\sigma,\sigma}|^{2}+k_{-\sigma}|t_{\sigma,-\sigma}|^{2}\right].
\end{equation}
Following the same procedure as above, the following expression
for the conductance in the parallel configuration is found
\begin{equation}\label{g-p}
    G_{P}\approx
    G_{P}^{\textrm{iso}}+G_{P}^{\textrm{aniso}}(\phi).
\end{equation}
The isotropic part is given by
\begin{equation}\label{giso-p}
    G_{P}^{\textrm{iso}}=G_{P}^{(0)}+G'_{P},
\end{equation}
where
\begin{equation}\label{g-0-p}
    G_{P}^{(0)}=\frac{e^2}{h}\sum_{\sigma=\uparrow,\downarrow}\left\langle \frac{4k_{F,\sigma}^{2}}
    {4k_{F,\sigma}^{2}+Q^{2}} \right\rangle_{\sigma},
\end{equation}
and
\begin{equation}\label{gprime-p}
    G'_{P}=-\frac{e^2}{2h}(\lambda_{\alpha}^{2}+\lambda_{\gamma}^{2})\sum_{\sigma=\uparrow,\downarrow}
    \langle g_{1\sigma}^{P}k_{\parallel}^{2} \rangle_{\sigma}.
\end{equation}
The anisotropic contribution reads
\begin{equation}\label{gani-p}
    G_{P}^{\textrm{aniso}}(\phi)=\frac{e^2}{h}\lambda_{\alpha}\lambda_{\gamma}\sum_{\sigma=\uparrow,\downarrow}
    \langle g_{2\sigma}^{P}k_{\parallel}^{2}
    \rangle_{\sigma}
    \cos(2\phi).
\end{equation}
The functions $g_{1\sigma}^{P}$ and $g_{2\sigma}^{P}$ are given by
\begin{equation}\label{g1-iso-p}
    g_{1\sigma}^{P}=\frac{8k_{F,\sigma}\left(2k_{F,\sigma}+k_{F,-\sigma}\right)Q^{2}-32
    k_{F,\sigma}^{3}k_{F,-\sigma}}{\left(4k_{F,\sigma}^{2}+Q^{2}\right)^{2}\left(4k_{F,-\sigma}^{2}+
    Q^{2}\right)}-g_{2\sigma}^{P},
\end{equation}
and
\begin{widetext}
\begin{equation}\label{g2-ani-p}
    g_{2\sigma}^{P}=\frac{4k_{F,\sigma}\left(k_{F,-\sigma}-k_{F,\sigma}\right)}
    {(4k_{F,\sigma}^{2}+Q^{2})^{3}(4k_{F,-\sigma}^{2}+
    Q^{2})}
    \left[16k_{F,-\sigma}k_{F,\sigma}^{3}-12k_{F,\sigma}\left(k_{F,\sigma}+k_{F,-\sigma}\right)Q^{2}
    +Q^{4}\right],
\end{equation}
\end{widetext}
respectively.

On the other hand, for the case of antiparallel configuration one
obtains the following expressions
\begin{equation}\label{g-ap}
    G_{AP}\approx
    G_{AP}^{\textrm{iso}}+G_{AP}^{\textrm{aniso}}(\phi),
\end{equation}
where
\begin{equation}\label{giso-ap}
    G_{AP}^{\textrm{iso}}=G_{AP}^{(0)}+G'_{AP},
\end{equation}
\begin{equation}\label{g-0-ap}
    G_{AP}^{(0)}=\frac{e^2}{h}\sum_{\sigma=\uparrow,\downarrow}\left\langle \frac{4k_{F,\sigma}
    \left(k_{F,-\sigma}-k_{F,\sigma}\right)}
    {\left(k_{F,-\sigma}+k_{F,\sigma}\right)^{2}+Q^{2}} \right\rangle_{\sigma},
\end{equation}
and
\begin{equation}\label{gprime-ap}
    G'_{AP}=-\frac{e^2}{2h}(\lambda_{\alpha}^{2}+\lambda_{\gamma}^{2})\sum_{\sigma=\uparrow,\downarrow}
    \langle g_{1\sigma}^{AP}k_{\parallel}^{2} \rangle_{\sigma}.
\end{equation}
The anisotropic contribution for the antiparallel case reads
\begin{equation}\label{gani-ap}
    G_{AP}^{\textrm{aniso}}(\phi)=\frac{e^2}{h}\lambda_{\alpha}\lambda_{\gamma}\sum_{\sigma=\uparrow,\downarrow}
    \langle g_{2\sigma}^{AP}k_{\parallel}^{2}
    \rangle_{\sigma}
    \cos(2\phi),
\end{equation}
and the functions $g_{1\sigma}^{AP}$ and $g_{2\sigma}^{AP}$ are
given by
\begin{eqnarray}\label{g1-iso-ap}
    &&g_{1\sigma}^{AP}=\frac{8k_{F,\sigma}\left(2k_{F,-\sigma}+
    k_{F,\sigma}\right)Q^{2}}{\left[\left(k_{F,\sigma}+
    k_{F,-\sigma}\right)^{2}+
    Q^{2}\right]^{3}}-g_{2\sigma}^{AP}\nonumber \\
    &&+\frac{8k_{F,\sigma}\left(k_{F,\sigma}-
    2k_{F,-\sigma}\right)\left(k_{F,\sigma}+
    k_{F,-\sigma}\right)^{2}}{\left[\left(k_{F,\sigma}+
    k_{F,-\sigma}\right)^{2}+
    Q^{2}\right]^{3}},
\end{eqnarray}
and
\begin{equation}\label{g2-ani-ap}
    g_{2\sigma}^{AP}=\frac{4k_{F,\sigma}\left(k_{F,\sigma}-k_{F,-\sigma}\right)}
    {\left[\left(k_{F,\sigma}+
    k_{F,-\sigma}\right)^{2}+
    Q^{2}\right]^{2}},
\end{equation}
respectively.

In the derivation of Eqs.~(\ref{gprime-p}), (\ref{gani-p}),
(\ref{gprime-ap}), and (\ref{gani-ap}) we assumed that the spin
orbit parameters $\bar{\alpha}$ and $\bar{\gamma}$ are independent
of the relative magnetization orientations in the ferromagnets.
Strictly speaking the values of $\bar{\alpha}$ and $\bar{\gamma}$
are $\theta$-dependent and therefore to the parallel and
antiparallel configurations correspond slightly different sets of
effective spin orbit parameters. This effect turns out to be
negligible for the system here investigated and we therefore omit
it.

In both the parallel and antiparallel cases, the averages have the
same meaning as in Eq.~(\ref{aver}) but now with
$k_{_{\textrm{max}}}^{\sigma}$ defined as
\begin{equation}\label{kmax-fsf}
    k_{_{\textrm{max}}}^{\sigma}=\sqrt{\frac{2m_{0}}{\hbar^{2}}\left(E_{F}+\sigma
    \frac{\Delta_{l}}{2}\right)}.
\end{equation}

It is worth noting that for a junction with structure inversion
symmetry the average Bychkov-Rashba parameter vanishes. One may
think that since we are considering a structure with leads of the
same material the anisotropic term in the conductance
corresponding to the parallel configuration (for the parallel
configuration the structure becomes symmetric under spatial
inversion) will vanishes. This is so for strictly symmetric under
spatial inversion structures. In practice, however, the interfaces
may not be identical, e.g., one of the interfaces may be epitaxial
while the other not. Another possibility is to consider different
terminations of the semiconductor at the left and right
interfaces. For the case of Fe/GaAs/Fe structures, for example,
one of the Fe/Gas interface may be Ga terminated while the other
may be As terminated.\cite{Erwin2002:PRB,Zega2006:PRL} This
interface-induced asymmetry is enough for the average
Bychkov-Rashba parameter to have a sizable value.

%\appendix{Appendix B}
\section{}
\label{appB}

Here we discuss the details about the solutions of the SSOM and
provide approximate expressions for the tunneling coefficients.

By requiring the probability flux conservation across the
interfaces one obtains that the eigenfunctions of the Hamiltonian
in Eq.~(\ref{hamilt-ssom}) should fulfil the following boundary
conditions
\begin{equation}\label{match-1}
    \Psi_{\sigma}^{(i)}(z_{ij})=\Psi_{\sigma}^{(j)}(z_{ij}),
\end{equation}
\begin{widetext}
\begin{eqnarray}\label{match-2}
    &&\left.\frac{\hbar^{2}}{2m_{i}}\left[1-
    \frac{2m_{i}\gamma_{i}}{\hbar}(k_{x}\sigma_{x}-k_{y}\sigma_{y})\right]
    \frac{d \Psi_{\sigma}^{(i)}}{dz}\right|_{z=z_{ij}}-
    \left.\frac{\hbar^{2}}{2m_{j}}\left[1-
    \frac{2m_{j}\gamma_{j}}{\hbar}(k_{x}\sigma_{x}-k_{y}\sigma_{y})\right]
    \frac{d\Psi_{\sigma}^{(j)}}{dz}\right|_{z=z_{ij}}=\nonumber \\
    &-&\alpha_{ij}(k_{x}\sigma_{y}-k_{y}\sigma_{x})\Psi_{\sigma}^{(i)}(z_{ij}),
\end{eqnarray}
\end{widetext}
where $i,j=l,c,r$ and $z_{lc}=0$ and $z_{cr}=d$ are the locations
of the left and right interfaces, respectively. The interface
Bychkov-Rashba parameters are introduced as
$\alpha_{lc}=\alpha_{l}$, $\alpha_{cr}=-\alpha_{r}$, while the
Dresselhaus parameter is $\gamma_{c}=\gamma$ in the semiconductor
and vanishes elsewhere (i.e., $\gamma_{l}=\gamma_{r}=0$).

Applying the above boundary conditions to the scattering states
given in Section \ref{ssom} one obtains a system of 8 linear
equations for determining the 8 unknown expansion coefficients.
The exact expressions for the transmission coefficients
$t_{\sigma,\sigma}$ and $t_{\sigma,-\sigma}$ are quite cumbersome.
However, simplified analytical expressions for $t_{\sigma,\sigma}$
and $t_{\sigma,-\sigma}$ are found in the limit $q_{0}d \gg 1$. In
such a case one finds the following approximate relations for the
tunneling coefficients,
\begin{equation}\label{tss-tsf}
    t_{\sigma,\sigma}=-\frac{D_{\sigma,\sigma}}{D};\;\;t_{\sigma,-\sigma}=\frac{D_{\sigma,-\sigma}}{D},
\end{equation}
where $D=f_{-}(-)f_{+}(-)-f_{-}(+)f_{+}(+)$, with
\begin{widetext}
\begin{equation}
%\begin{eqnarray}
\label{efes}
    %&&
    f_{\pm}(\lambda)=\pm
    \frac{i\alpha_{r}k_{\parallel}Q}{2 V_{0}}\left(\sigma e^{i(\phi_{r}-\xi+\pi/4)}+\lambda
    e^{2i\xi}\right)
    %\nonumber \\
    %&&
    +\frac{d}{2}\left(\frac{m_{0}}{m_{\pm \lambda}}q_{\mp \lambda}-i\kappa_{\pm
    \sigma}\right)\left(1-\lambda \sigma
    e^{i(\phi_{r}+\xi+\pi/4)}\right).
%\end{eqnarray}
\end{equation}
\end{widetext}
and
\begin{equation}\label{mass-pm}
    \frac{1}{m_{\pm}}=\frac{1}{m_{c}}\left(1 \pm \frac{2m_{c}\gamma
    k_{\parallel}}{\hbar^{2}}\right).
\end{equation}
Furthermore, we have
\begin{equation}\label{dss}
    D_{\sigma,\sigma}=\frac{2m_{0}d}{m_{+}}q_{-}f_{-}(+)g_{-}-\frac{2m_{0}d}{m_{-}}q_{+}f_{-}(-)g_{+},
\end{equation}
and
\begin{equation}\label{dsf}
    D_{\sigma,-\sigma}=\frac{2m_{0}d}{m_{+}}q_{-}f_{+}(-)g_{-}-\frac{2m_{0}d}{m_{-}}q_{+}f_{+}(+)g_{+}.
\end{equation}
In these equations we introduced the notation
\begin{widetext}
\begin{equation}\label{gees}
    g_{\pm}=\frac{id\sqrt{k_{\sigma}}\left[\left(f_{0}\mp h_{1}-\frac{m_{0}d}{m_{\pm}}q_{\mp}\right)
    \left(1\pm \sigma e^{i(\phi_{l}+\xi+\pi/4)}\right)\mp h_{2}\left(1\mp \sigma e^{i(\phi_{l}+\xi+\pi/4)}\right)
    \right]e^{-q_{\pm}d}}
    {h_{2}^{2}+\left(f_{0}-h_{1}-\frac{m_{0}d}{m_{+}}q_{-}\right)\left(f_{0}+h_{1}-\frac{m_{0}d}{m_{-}}q_{+}\right)},
\end{equation}
\end{widetext}
where $f_{0}=i(k_{\sigma}+k_{-\sigma})d/2$,
\begin{equation}\label{h1}
    h_{1}=\frac{i\sigma
    d}{2}\left(k_{\sigma}-k_{-\sigma}\right)\cos(\phi_{l}+\xi+\pi/4)+\frac{\alpha_{l}k_{\parallel}Q}{V_{0}}\sin(2\xi),
\end{equation}
and
\begin{equation}\label{h2}
    h_{2}=-\frac{\sigma
    d}{2}\left(k_{\sigma}-k_{-\sigma}\right)\sin(\phi_{l}+\xi+\pi/4)-i\frac{\alpha_{l}k_{\parallel}Q}{V_{0}}\cos(2\xi).
\end{equation}

It is worth noting that the approximate expressions for the
tunneling coefficients here provided are valid up to first order
in $\exp(-q_{\pm}d)$. This approximation is appropriate for
treating junctions with high and not too thin potential barriers.
For the systems here considered the hight of the barrier (with
respect to the Fermi level) is about $V_{b}=(V_{0}-E_{F})\approx
0.75 \textrm{ eV}$ and $d$ varies from $20 \textrm{ \AA}$ to $200
\textrm{ \AA}$. In such cases the approximations here discussed
turns out to be excellent.

It is not difficult to show that in the limit
$\alpha_{l}=\alpha_{r}=\gamma=0$, the expressions for the
tunneling coefficients here obtained reduce to the ones reported
in Ref.~\onlinecite{Fabian2007:APS} for the case of vanishing SOI.

We also remark that the expressions above were obtained for the
general case of a F/S/F tunnel junction but the corresponding
expressions for a F/S/NM junction can easily be obtained by taking
the limits $\kappa_{\sigma}=\kappa_{-\sigma}=\kappa$ and
$\phi_{l}=\phi_{r}=\phi$.

%\bibliographystyle{apsprb}

%\bibliography{refs}

\begin{thebibliography}{47}
\bibitem[{not({\natexlab{b}})}]{note2}
\bibinfo{note}{Note also that when the effective Bychkov-Rashba parameter
  $\bar{\alpha}$ changes sign, the axes of symmetry of $|\mathbf{w}|$ are
  flipped by $90^{\circ}$ and the situation above explained is reversed (i.e.,
  now the transmission corresponding to the magnetization direction [110]
  becomes the dominant one), leading to the inversion of the TAMR effect.}
\end{thebibliography}

% \end{multicols}

\end{document}